\documentclass[usenatbib]{mn2e}
\usepackage{amsgen,amsmath,amstext,amsbsy,amsopn,amssymb,amsfonts,amsmath}
\usepackage{graphicx}
\usepackage[usenames]{color}

\title[Large scale bias]
{Large scale bias and the inaccuracy of the peak-background split}

\author[M. Manera, R. K. Sheth, R. Scoccimarro]
{M. Manera$^{1}$\thanks{E-mail:  manera@nyu.edu}
Ravi K. Sheth$^{2}$ \thanks{E-mail:  shethrk@physics.upenn.edu}
\& R. Scoccimarro$^1$\thanks{E-mail: rs123@nyu.edu}
\footnotemark[1]\\
$^{1}$Center for Cosmology and Particle Physics, 
     Department of Physics, New York University, 
     New York, NY 10003, USA\\
$^{2}$Center for Particle Cosmology, University of Pennsylvania, 
     209 S. 33 St., Philadelphia, PA 19104, USA}

\newcommand{\bm}[1]{{\mbox{\boldmath $#1$}}}

\newcommand{\beq}{\begin{equation}}
\newcommand{\eeq}{\end{equation}}
\newcommand{\beqa}{\begin{eqnarray}}
\newcommand{\eeqa}{\end{eqnarray}}

\newcommand{\lexp}{\mathop{\langle}}
\newcommand{\rexp}{\mathop{\rangle}}
\newcommand{\rexpc}{\mathop{\rangle_c}}

\def\d{\delta}

\begin{document}

\maketitle

\label{firstpage}

\begin{abstract}
The peak-background split argument is commonly used to relate the 
abundance of dark matter halos to their spatial clustering.  
Testing this argument requires an accurate determination of the 
halo mass function.  We present a Maximum Likelihood method for 
fitting parametric functional forms to halo abundances which differs 
from previous work because it does not require binned counts.  
Our conclusions do not depend on whether we use our method or more 
conventional ones.  In addition, halo abundances depend on how halos 
are defined.  Our conclusions do not depend on the choice of link 
length associated with the friends-of-friends halo-finder, nor 
do they change if we identify halos using a spherical overdensity 
algorithm instead. 
The large scale halo bias measured from the matter-halo cross spectrum 
$b_\times$ and the halo autocorrelation function $b_\xi$ 
(on scales $k\sim 0.03h\,{\rm Mpc}^{-1}$ and $r\sim 50h^{-1}$Mpc) 
can differ by as much as 5\% for halos that are significantly 
more massive than the characteristic mass $M_*$.  
At these large masses, the peak background split estimate of the 
linear bias factor $b_1$ is 3-5\% smaller than $b_\xi$, which is 
5\% smaller than $b_\times$. 
We discuss the origin of these discrepancies:  deterministic nonlinear 
local bias, with parameters determined by the peak-background split 
argument, is unable to account for the discrepancies we see.  
A simple linear but nonlocal bias model, motivated by peaks theory, 
may also be difficult to reconcile with our measurements.  More work 
on such nonlocal bias models may be needed to understand the nature 
of halo bias at this level of precision.
\end{abstract}


\section{Introduction}
Halo abundances and clustering are both crucial ingredients in 
the halo model of large scale structure \citep{Peacock:2000qk, 
Seljak:2000gq, Scoccimarro:2001howmany,Cooray:2002HaloModel}.  
However, following \citet{ST99}, the two are not indepedendent:  
an accurate model of halo clustering is part and parcel of an 
accurate model of halo abundances.  This is because of an argument 
that has come to be called the peak-background split 
\citep{BBKS:1986,CK:1989,MW:1996}, in which, on large scales, 
perturbed regions of the matter field are treated as though they 
are universes with slightly different mean density and Hubble 
constant \citep[for an explicit calculation, see][]{MS:2009}.  

As a result, there has been considerable effort to provide simple, 
accurate and physically motivated functional forms for the halo 
mass function \citep{PS74,Bond:1991,Lee1998,Sheth:2001elipsoidal}, and to 
determine if such models provide adequate descriptions of the 
simulations.  When appropriately scaled, the functional form predicted 
by \citet{PS74} is independent of power spectrum and cosmology.  
\citet{ST99} showed that, although this sort of rescaling of the 
mass function is not expected to hold exactly for the CDM family of 
models, it does produce an approximately universal curve in 
simulations, although the functional form of this universal curve is 
different from that of \citet{PS74}.  Subsequent work has confirmed 
that the mass function is indeed approximately universal 
\citep{Jenkins:2001,Reed:2003}, 
with only the most recent measurements beginning to detect the 
expected departures from universality 
\citep{White2002,Reed:2007,Tinker:2008}.  
This is simply because the departures are small so large simulation 
volumes are required to see the effect with high significance. 

The main goal of the present paper is to use the more precise 
measurements of halo abundances which can now be made (in simulations) 
to perform more precise tests of how well the peak background split 
argument works.  We do so by measuring halo abundances and clustering 
in large volumes, and then comparing the clustering signal with that 
predicted from the measured abundances by the peak background split 
ansatz.  
We assess the robustness of our results by varying how we identify 
halos in the simulations; in each case, we use two different 
parametrizations for our measured abundances, and three different 
methods for fitting the parametrized models to the measurements.  
We then compare the predicted and measured clustering signals 
in both real and Fourier space, and we do all this for two 
(and sometimes three) different redshifts.  

At this level of precision, the comparison of measurement and 
prediction is somewhat subtle, because it depends on the details 
of whether or not the bias is expected to be deterministic or 
stochastic, local or nonlocal, linear or nonlinear, constant or 
scale-dependent.  We study two limiting cases in detail: 
a bias which is deterministic and local in configuration space, 
and is scale independent at linear order but contains higher 
order nonlinear terms, and a bias which is deterministic and linear 
in Fourier space, with no higher order terms, but the linear bias 
is $k$-dependent.  The former arises naturally in the simplest 
models of halo abundances; the latter is motivated by associating 
nonlinear stuctures with peaks in the initial density fluctuation 
field.  

This paper is organized as follows: 
Section~\ref{bkgnd} gives some theoretical background and describes a 
number of ways one might have quantified the bias between the halo and 
matter distributions.  It then specifies the particular ways we have 
adopted for our test.  
Section~\ref{sims} presents measurements of halo abundances and 
clustering in our simulations, and comparison with the bias 
predicted by the peak background split argument.  
A final section summarizes our results and conclusions. 
Appendix A describes a number of ways we have attempted to fit the 
halo mass function, one of which is a new Maximum Likelihood
estimator of halo abundances that does not require binned counts.
Appendix B provides explicit expressions for the peak background 
split bias factors associated with our parametrizations of the 
halo mass function.  

\section{Background}\label{bkgnd}
\subsection{Counts in cells and the peak background split}
The peak background split \citep{BBKS:1986,CK:1989} is an approximation 
in which the effect of long wavelength density perturbations on 
structure formation is simply to modify the collapse times of 
non-linear objects.  This modification depends on the density of 
the perturbed region but not on its volume.  It is common to state 
that the number density of halos in a perturbed region is expected 
to be the same as that of an unperturbed region, but at a slightly 
different time.  However, it is better to think of the perturbed 
number density as being the same as that of an unperturbed region 
in a different background cosmology (after all the density is 
different), but one that has the same age (meaning the effective 
Hubble constant is different) \citep{MS:2009}.  When expressed in 
terms of linear theory quantities, this effect changes the critical 
density for non-linear collapse in a way that depends on the 
nonlinear density of the perturbation \citep{MW:1996}.  

Thus, while in general the mean number of halos of mass $m$ in a 
cell depends on its volume $V$ and mass $M$, in this approximation, 
for cells for cells which are sufficiently large that $m\ll M$,
the overdensity of halos depends, not on $M$ and $V$, but on $M/V=1+\delta$.  
That is to say, 
\begin{equation}
\langle N_{\rm h}(m,\delta_c|M,V)\rangle \equiv
n_{\rm h}(m,\delta_c)V\, [1 + \langle\delta_h(m|\delta)\rangle]
\end{equation}
where $n(m,\delta_c)$ is the average number density of halos with 
mass $m$, and 
\begin{equation}
\langle\delta_h(m)|\delta\rangle = \sum_{k>0} \frac{b_k(m,\delta_c)}{k!}\,
             \Bigl(\delta^k - \langle\delta^k\rangle\Bigl).  
\label{localb}
\end{equation}
The coefficients $b_k(m,\delta_c)$ come from Taylor expanding 
$n(m,\delta_c-\delta)$ around $\delta=0$, and the $\langle\delta^k\rangle$ 
terms are required if one wishes to truncate the expansion at finite 
$k$ but still enforce $\langle\delta_h(m)|\delta\rangle=0$.  
Thus, in this framework, halo bias is {\em deterministic} ($\delta$ is 
the only random field that determines $\delta_h$) but {\em nonlinear} 
(high order terms in $\delta$ contribute), so it is of the form 
discussed by e.g. \cite{FG93}.  

The most direct check of this assumption is to measure the quantity 
on the left hand side of equation~(\ref{localb}) in large cells $V$, 
and compare with the coefficients one predicts from the mass function
\citep{SL99,S307}.  Note that this is explicitly a real-space, 
counts-in-cells calculation.  
It is, however, a difficult approach, since the halo bias 
coefficients of interest are those for large cells, but these tend 
to have small variance (the universe is homogeneous on large scales), 
meaning that there is only a small range of $\delta$ over which to 
measure the shape of the halo bias relation.  In practice, measuring 
$b_2$ is tough, and $b_3$ is even more challenging.  

\subsection{Other measures of the linear bias factor}
A less direct measure of this bias is given by the volume average 
of the cross correlation function between halos and mass.  In this 
case, one measures 
\begin{eqnarray}
1 + \sigma^2_{\rm hm}(V) &=& \int dM \,p(M|V)\,\sum p(N_h|M,V)
                          \frac{M}{\bar\rho V}\,\frac{N_h}{n_{\rm h}V}
                          \nonumber\\
  &=& \int dM \,p(M|V)\,\frac{M}{\bar\rho V}\,
              \frac{\langle N_h|M,V\rangle}{n_{\rm h}V} \nonumber\\
  &=& 1 + \sum_{k>0} \frac{b_k}{k!} \langle\delta_M^{k+1}\rangle \nonumber\\
  &=& 1 + b_1 \sigma_M^2 + \ldots
\end{eqnarray}
where $\sigma^2_{\rm hm}(V)$ is the cross-correlation between halo and
mass counts in cells of size $V$, $p(M|V)$ is the probability a randomly
chosen cell of size $V$ contains mass $M$, and 
\begin{equation}
\sigma_M^2 \equiv \langle\delta_M^2\rangle 
 = \int \frac{dk}{k}\,\frac{k^3\,P(k)}{2\pi^2}\,W^2(kR)
\label{sigmaM}
\end{equation}
where $P(k)$ is the power spectrum of the mass, and $W$ is the 
Fourier transform of the smoothing volume (so $V\propto R^3$).  

And even more indirect is the second factorial moment of the halo 
counts-in-cells:  
\begin{eqnarray}
1 + \sigma^2_{\rm hh}(V) &=& \int dM \,p(M|V)\,\sum p(N_h|M,V)
                          \frac{N_h}{n_{\rm h}V}\frac{N_h-1}{n_{\rm h}V}
                          \nonumber\\
  &=& \int dM \,p(M|V)\,
           \frac{\langle N_h(N_h-1)|M,V\rangle}{(n_{\rm h}V)^2}.
\end{eqnarray}
If the halo counts in cells $(M,V)$ follow a Poisson distribution 
around the mean $\langle N_h|M,V\rangle$
(this is a bad assumption when $m$ is not small compared to $M$), 
then this becomes 
\begin{eqnarray}
1 + \sigma^2_{\rm hh}(V) &=& \int dM \,p(M|V)\,
           \frac{\langle N_h|M,V\rangle^2}{(n_{\rm h}V)^2} \nonumber\\
  &=& 1 + b_1^2 \sigma_M^2 + \ldots 
\end{eqnarray}
Finally, it is worth noting that 
\begin{eqnarray}
\sigma^2_{\rm hm}(R) &=& 
\int \frac{dk}{k}\, \frac{k^3P_{\rm hm}(k)}{2\pi^2}\, W^2(kR)\\
   &=& 4\pi \int_0^{2R} dr r^2\,\xi_{\rm hm}(r)\,
            \frac{3}{\pi}\frac{(4 + r/R)(2- r/R)^2}{32R^3} \nonumber
\end{eqnarray}
where the final expression assumes tophat smoothing.  
Similar relations hold for $\sigma_{\rm hh}$, $\xi_{\rm hh}$ and $P_{\rm hh}$.

So, if $b_1$ is independent of scale, then the slope of the regression 
of $\delta_h$ on $\delta_m$ is the same quantity as 
$\sigma_{\rm hm}^2/\sigma^2$ and $\xi_{\rm hm}/\xi$; 
and if the counts are Poisson, then this is also the same as
$\sqrt{\sigma^2_{\rm hh}/\sigma^2}$, $\sqrt{\xi_{\rm hh}/\xi_{\rm dm}}$, 
$\sigma^2_{\rm hh}/\sigma^2_{\rm hm}$, and $\xi_{\rm hh}/\xi_{\rm hm}$
at large scales.
In addition, if $b_1$ is independent of scale, then the bias in 
Fourier space quantities is simply related to (equal to!) those 
in configuration space.  In particular, 
$\sqrt{P_{\rm hh}(k)/P(k)}$, $P_{\rm hh}(k)/P_{\rm hm}(k)$ 
and $P_{\rm hm}(k)/P(k)$ should all equal $b_1$ at low k.  
But in general, all these quantities are different.  
We discuss some of the differences expected in concrete bias models 
and in view of our measurements below.

Even if these bias factors are equal, actually estimating $P_{\rm hh}$ 
is difficult because the measurement requires a shot-noise correction 
for the discreteness of the halos.  Because the massive halos of most 
interest in the present study are rare, this correction can be 
significant, but because they are strongly clustered, this correction 
is currently uncertain \citep{S307}. 
There is no shot-noise correction for $P_{\rm hm}$, so, in what follows, 
this is the statistic we will use to test the peak background split 
expression for the linear bias parameter $b_1$.  We also test the 
ratio $\sqrt{\xi_{\rm hh}/\xi_{\rm dm}}$, for which no shot-noise 
correction is necessary.  

\subsection{The effects of nonlinearity on large-scale bias}\label{bnl}
Differences between the predicted $b_1$ and the large scale bias 
measured from correlation functions are expected if the bias is 
nonlinear.  Indeed, the peak-background split itself predicts that 
halo bias is not linear (the higher order coefficients in 
equation~\ref{localb} are generically non-zero), and such 
nonlinearities are seen in numerical simulations \cite[see, e.g., 
scatter plots of $\delta_h$ vs $\delta_m$ in Appendix~B of][]{Smith:2007}.  
This complicates interpretation of the measured values of $P_{hm}/P_{mm}$ 
and $\sqrt{\xi_{\rm hh}/\xi_{\rm dm}}$ as follows.  

In the local bias framework of equation~(\ref{localb}), the halo-mass 
cross-correlation reads
\beq
\lexp \d_{h1} \d_2 \rexp = 
 b_1 \lexp \d_1 \d_2 \rexp + {b_2\over 2} \lexp \d_1 \d_2^2 \rexp 
 + {b_3\over 6} \lexp \d_1 \d_2^3 \rexp + \ldots
\label{cross-corr}
\eeq
where 1 and 2 denote two different spatial positions.  
In the large-scale limit, 
perturbation theory says that 
\beq
\lexp \d_1^p \d_2^q \rexpc \equiv C_{pq}\, \sigma^{2(p+q-2)}_R\ \xi
\label{Cpq}
\eeq
where $\sigma^2_R$ denotes the variance in the dark matter field 
when smoothed on scale $R$, and $C_{pq}$ are closely related to 
the skewness, kurtosis and so on.  E.g.,
$C_{21} = 68/21+\gamma_R/3$, with
$\gamma_R \equiv d\ln \sigma^2_R/d\ln R$ and
$C_{pq}=C_{p1}C_{q1}$ \citep{Bernardeau96,GaztanagaFosalbaCroft02}.  
Thus, on large scales, the cross-correlation bias is 
\beq
b_\times \equiv {\lexp \d_{h1} \d_2 \rexp \over \xi}
         = b_1 + {\sigma^2_R\over 2} \Big( C_{21}\, b_2 + b_3\Big) 
               + {\sigma^4_R\over 6} C_{31}\, b_3 + \ldots ,
\label{bcross}
\eeq
and it applies equally well in configuration and Fourier space.  
Keeping only the first order corrections to linear bias, yields 
\begin{equation}
b_\times = \frac{P_{hm}(k|R)}{P(k)}
         = b_1 + \sigma_R^2 
\left[ \Big(\frac{34}{21}+{\gamma_R \over 6}\Big) b_2 + \frac{b_3}{2} \right] 
\label{nonlinearbias}
\end{equation}
for the Fourier-space quantity 
\citep[e.g.][who neglected the $\gamma_R$ term]{Smith:2007}, 
where $P_{hm}(k|R)$ denotes the cross-power of the halo and mass 
fields when both have been smoothed with a filter of scale $R$.

In the present context, for halos of a given mass, the peak-background 
split argument gives the values of $b_i$.  However, the choice of 
smoothing scale $R$ is less straightforward.  It must be large enough 
that the assumptions of a deterministic, scale independent bias are 
reasonably accurate, so $R$ must be substantially larger than the 
Lagrangian radius of the halos \citep{SL99, S307, ManeraGazta09}.  
But there is no other underlying theory for this scale.  

The same logic that led to equation~(\ref{bcross}) says that 
\beqa
b^2_\xi &=&  \frac{\lexp \d_{h1} \d_{h2} \rexp}{\xi}
          =    b_1^2 + b_1\,\sigma^2_R\, (C_{21}\, b_2 + b_3) 
                     +  {b_2^2\over 2} \, \xi \nonumber \\
         & & \qquad\qquad\qquad + \sigma^4_R \, 
  \Big({b_1 b_3\over 3}\, C_{13} + {b_2^2\over 4}\, C_{22} \Big)+ \ldots
            \nonumber\\
&\simeq& b_\times^2 - \sigma^4_R \,\frac{b_3}{4}
         \left(b_3 + 2 b_2 C_{12} \right)\, + \frac{b_2^2}{2}\, \xi.
\label{bauto}
\eeqa
The final expression shows that $b_\xi\ne b_\times$ even when 
$\sigma^2_R \ll 1$.  
And the $\xi$ term in $b_\xi$ generates a shot-noise contribution 
at low-$k$ in the power spectrum.  

\begin{table}
\begin{tabular}{c p{4.3cm} c} 
bias symbol  & meaning  & equation \\
\hline 
$b_1$, $b_2$, $b_3$ & First (linear), second and third order bias from 
                     Taylor expansion of the fluctuation in 
                     the mass density field.  This is a deterministic 
                     local bias model for which predictions exist 
                     from the peak background split argument in the 
                     large cell limit. &  (2) \\
$b_\times$ & Large scale bias from the matter-halo cross power. 
            Values taken at $k=0.03 h$Mpc$^{-1}$. & (11)  \\
$b_\xi$ & Large scale bias from the correlation function. 
         Values are taken by averaging $\xi$ over 
         $40\le r\le 60 h^{-1}$Mpc. & (12)\\
$b_\nu$, $b_\zeta$  & 
         Linear and quadratic bias from the high peaks model. & (13)   \\
\hline  
\end{tabular}
\caption{Notation for the various bias factors used in this paper.} 
\label{notationtable}
\end{table}

\subsection{The peaks-bias model}
The previous discussion supposed that the fundamental quantity was 
the bias between halo and mass counts in cells.  An alternative model 
is that (high) peaks in the initial density field are the seeds around 
which massive halos form \citep{Kaiser1984}.  
In this case the large scale bias is simplest in Fourier space:  
\begin{equation}
\delta_{pk}(k) = (b_\nu + b_\zeta k^2)\,W_{pk}(kR_{pk})\delta(k), 
\label{deltapk}
\end{equation}
where $W_{pk}$ is the smoothing filter with which the peak was 
identified \citep{Matsubara1999,Desjacques2008}.  

Typically, to approximate halos of mass $m$ by peaks, one uses a 
Gaussian smoothing filter with $m \propto R_{pk}^3$.  In this case, 
a halo of mass $m$ is associated with a peak of height 
$\nu = \delta_{pk}/\sigma_0$, where $\delta_{pk}$ is of order unity 
as suggested by the spherical evolution model, and $\sigma_0^2$ is 
given by equation~(\ref{sigmaM}) but with smoothing scale $R_{pk}$.  
At high masses, the resulting peak mass function is similar to that 
of halos \citep{Sheth:2001peaks}.  The quantity 
 $b_\zeta\propto (\sigma_0/\sigma_1)^2\,(\nu/\sigma_0 - b_\nu)$, 
where $\sigma_1^2$ is similar to $\sigma_0^2$, but with an
extra factor of $k^2$ in the integral in equation~(\ref{sigmaM}).  
For a power law power spectrum with $P(k)\propto k^n$, 
$(\sigma_0/\sigma_1)^2\propto m^{2/3}$.  In the high peak 
($\nu\gg 1$) limit, $b_\nu\to (\nu - 3/\nu)/\sigma_0$ so 
$\nu/\sigma_0 - b_\nu \to 3/(\sigma_0\nu)$.  
In this limit, $b_\zeta$ increases as $m$ increases, and 
$(b_\zeta/b_\nu) \to (\sigma_0/\sigma_1)^2 (3/\nu^2) 
                \propto m^{2/3 - (n+3)/3}$, 
a point to which we will return later.

Equation~(\ref{deltapk}) implies that 
\begin{eqnarray}
\label{Ppk}
P_{pk,\delta}(k) &=& (b_\nu + b_\zeta k^2)\,W_{pk}(kR_{pk})\,P_{\rm L}(k),\\ 
P_{pk,pk}(k)     &=& (b_\nu + b_\zeta k^2)^2\,W^2_{pk}(kR_{pk})\,P_{\rm L}(k),
\label{Ppkpk}
\end{eqnarray} 
so $P_{pk,\delta}(k)/P(k)$, $\sqrt{P_{pk,pk}(k)/P(k)}$ and 
$P_{pk,pk}(k)/P_{pk,\delta}(k)$ all measure the same quantity 
(even though the quantity depends on $k$!), but the bias relations 
from correlation functions or counts in cells will be more complicated 
(because of the $k$ dependence).  In particular, notice that, in 
contrast to the previous model, here the linear bias factor itself 
is scale-dependent.  

Now, the bias relations above are for peaks identified in the 
initial fluctuation field. 
At this time $b_1$ from the peak background split calculation 
equals $b_\nu$ from the Fourier bias calculation \citep{DS09}.  
(In principle, at least for peaks, this agreement can be used as a guide 
to the appropriate shot-noise correction for $P_{pk,pk}(k)$ -- like 
massive halos, high peaks are rare, so the shot-noise correction 
matters -- but this is beyond the scope of this paper.)
A peak background split estimate for the late time bias parameters 
$b_1$, $b_2$, etc. of peaks was made by \cite{MJW97}.  This estimate 
says that $b_1\to 1+b_1$ (with similar consequences for $b_2$ etc.), 
and is in reasonable agreement with measurements in simulations of 
$\sqrt{\sigma^2_{pk,pk}/\sigma^2}$ and $\sqrt{\xi_{pk,pk}/\xi}$
\citep{MJW97} (i.e., within the accuracy of what was possible with 
the smaller simulation volumes of 10 years ago).  
This suggests that $b_\nu$ evolves as $b_1$, but a good model for 
the evolution of $b_\zeta$ is still not available.
Therefore, when we compare the peaks model with measurements in 
simultions, we will simply consider if a $k^2$ scaling of the bias 
factor seems appropriate, and if the onset of this term occurs at 
smaller $k$ for halos of higher masses.  

\section{Measurements in simulations}\label{sims}

\subsection{Description of the simulations}
For our analysis we use 49 cosmological dark matter simulations 
of a flat $\Lambda$CDM cosmology with $\Omega_m=0.27$, $\Omega_\Lambda=0.73$, 
$\Omega_b=0.046$, $\sigma_8=0.9$, $h=0.72$  and $n_s=1.0$.  
Each simulation was run using periodic boundary conditions in a box 
of size $L_{box}=1280h^{-1}$Mpc, which contains $640^3$ particles.
This gives a particle mass of $M_p \simeq 6\times 10^{11}h^{-1}M_\odot$.
All 49 runs have the same parameters except for the random seeds 
used to generate the initical conditions. Therefore they can be 
considered as different realizations (or parts) of the same universe; 
this allows us to estimate errors on the mass function and bias factors 
we measure in the next section.  
For reference, the total volume sampled by our runs is
$V_{\rm T} \simeq 102 h^{-3}$  Gpc$^3$.  

\begin{figure}
\includegraphics[width=3.5in]{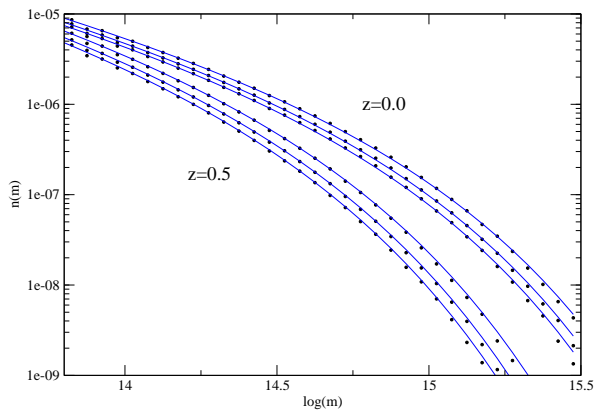}
\caption{Mass function at $z=0$ (upper set of curves) 
and $z=0.5$ (lower set of curves) for three linking lengths 
in simulations: 
0.15 (fewest massive halos), 0.168 and 0.2 (most massive halos). 
Lines show equation~(\ref{vfvst}) with parameters from our new 
Maximum Likelihood estimator (see Table~\ref{massfitSTtable}).}
\label{massfuncfig}
\end{figure}

\begin{figure}
\includegraphics[width=3.5in]{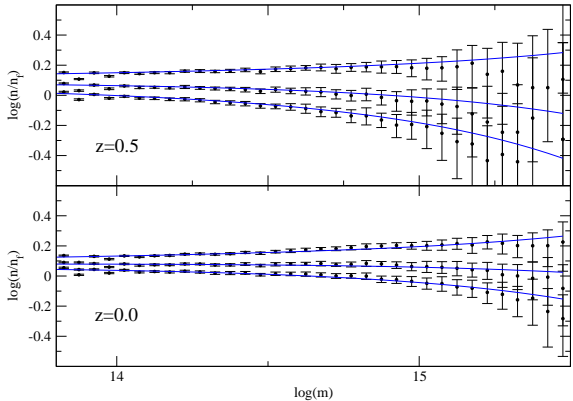}
\caption{Same as figure \ref{massfuncfig}, only now, to better see 
 the range on the plot, the mass functions have been divided 
 by a fiducial function (equation~\ref{vfvst} with $p=0.33$ and $q=0.75$). Error bars show
the rms variation between simulations.}
\label{massfuncfid}
\end{figure}

One potentially important difference from almost all previous work 
in which volumes of this size have been studied is in how we 
generate our initial conditions.  These are set at $z=50$ by 
using CMBFAST \citep{cmbfast} to generate the 
Transfer function for the initial matter power spectrum.  We then 
use a Second Order Lagrangian Perturbation Theory (2LPT) code 
\citep{Roman:1998} to generate the initial displacement field.  
The use of 2LPT initial conditions ensures that spurious transient 
effects in the simulations are negligible at low redshifts 
\citep{Crocce:2006}. 
The tree-PM code {\sc Gadget-2} \citep{gadget2}, with a softening 
length set to 20$h^{-1}$kpc, is then used to simulate the subsequent 
evolution.  

\subsection{The halo mass function}

We have run a standard friends-of-friends (FoF) code to identify 
dark matter halos in the simulations at redshifts $z=0$ and $z=0.5$. 
The halo mass function one obtains depends on the one free parameter 
of the FOF algorithm:  the linking length. 
Shorter linking lengths return lower mass halos.  
Since halo abundances and clustering strength are intimately related, 
the choice of linking length also affects the halo bias parameters. 
To address this, we have explored three choices:
$l_{\rm link}=0.15, 0.168$ and 0.2 (in units of the interparticle 
separation).

The halo mass of each object found by the FoF algorithm was determined 
from the number of particles $N$ it contains, corrected for discreteness 
effects following \cite{Warren:2006}. Thus, 
$M_h = M_p N_{\rm corrected}$, where $N_{\rm corrected}=N(1-N^{-0.6})$.  
This correction has been tested only for FoF halos with $l_{link}=0.2$,
and may sligtly overcorrect the mass for smaller linking lengths. Since
in this paper we are fitting the mass function for halos having more 
than 105 particles, these differences are negligible for the large 
mass halos which are of most interest in what follows.

It is common to use the same linking length for all redshifts.  
However, the natural outcome of the spherical collapse model 
predicts that, in $\Lambda$CDM models, halos are a larger 
multiple of the background density at late times.  
If this model is correct, then one expects the appropriate link 
length to be approximately constant at early times, and to 
decrease at late times.  Our choices of linking-length approximately 
bracket the expected range of densities.  

Another popular choice for identifying halos is to require them 
to be a fixed multiple of the critical density. In $\Lambda$CDM 
models, this has the virtue of being well-motivated at early times 
(when the background cosmology is effectively Einstein-de Sitter, so 
the background and critical densities are equal) as well as at very 
late times (when the critical density has become constant).  
In section \ref{sectionSO} we use halos identified using a spherical 
overdensity method by \cite{Tinker:2008}.  However, in this case, 
the overdensity was a fixed multiple (200) of the background density.  
We find that the main results which follow are robust to which 
halo finder we use.

\begin{figure}
\includegraphics[width=3.5in]{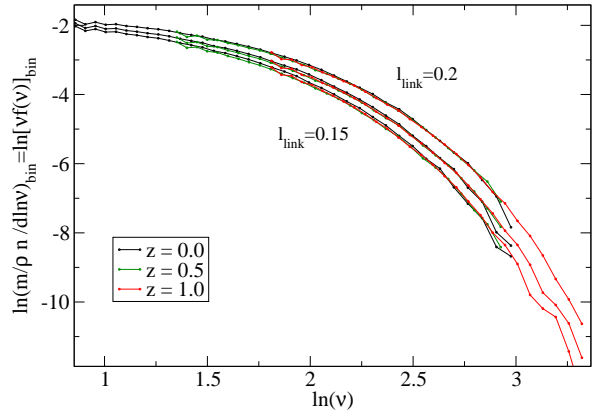}
\caption{Same as Figure~\ref{massfuncfig}, but now shown in scaled 
units, so outputs from $z=0, 0.5$ and $1$ are shown together.  
Because we only count halos with more than 105 particles, the 
lower redshift output probes to smaller $\nu$, and the higher 
redshift output to higher $\nu$. 
Results for the three linking lengths are shown: 0.15, 0.168 and 0.2. 
For a fixed $\nu$ larger $l_{\rm link}$ yields more halos.}
\label{vfv}
\end{figure}

\begin{figure}
\includegraphics[width=3.5in]{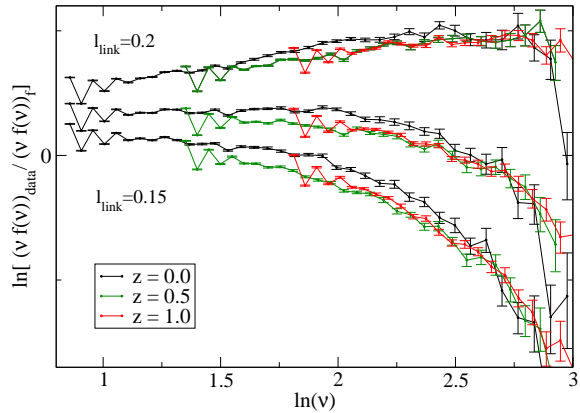}
\caption{Same as Figure~\ref{massfuncfid}, but now in scaled units. Error bars
show the error on the mean value between simulations.}
\label{vfvfid}
\end{figure}

Figure~\ref{massfuncfig} shows the mass functions associated with 
the three linking lengths at $z=0$ and $z=0.5$.  
To emphasize detailed differences, we show this same information 
divided by a fiducial model for halo abundances in 
Figure~\ref{massfuncfid}. The fiducial model is that 
of equation~(\ref{vfvst}) below, with $p=0.75$ and $q=0.33$.  
In these, as in all the plots to follow, the bins are 0.05~dex in mass, 
and error bars, unless stated otherwise, show the rms variation between 
simulations.  
The true error on the mean is a factor of $\sqrt{49}=7$ smaller. 
It is interesting to ask if the halo catalog returned by a shorter 
link-length is essentially a higher redshift version of the halo 
catalog associated with the longer link-length.  We will have more 
to say about this shortly, but note that this dependence on linking 
length is not naturally included in models of halo abundances 
\citep[e.g.][]{Sheth:2001elipsoidal}.  

When the masses are suitably rescaled, the mass function can be 
expressed in a functional form that is nearly universal - being 
approximately independent of time, cosmology, and initial power 
spectrum \citep{ST99}.  The spherical evolution model suggests that 
the natural scaling variable should be
\begin{equation}
\nu \equiv \frac{\delta^2_{\rm sc}(\Omega_z,\Lambda_z)}{D^2(z)\,\sigma_0^2(m)}
\label{definenu}
\end{equation}
where $\delta_{\rm sc}$ is the critical density required for 
spherical collapse in a cosmology with parameters $(\Omega_z,\Lambda_z)$, 
$D(z)$ is the linear theory growth factor in units of its value at 
$z=0$ [e.g. $D(z) = (1+z)^{-1}$ and $\delta_{\rm sc}(z)=1.686$ if 
$(\Omega_z,\Lambda_z)=(1,0)$], 
and 
\begin{equation}
\sigma_0^2(m) = \int \frac{dk}{k}\,\frac{k^3\,P_0(k)}{2\pi^2}\,W^2(kR_m)
\label{sigma0}
\end{equation}
with $m = \bar\rho\, (4\pi R_m^3/3)$ and 
$W(x) = (3/x^3)\,(\sin x - x\cos x)$.  Here $P_0(k)$ denotes the 
initial power spectrum of fluctuations, scaled using linear theory 
to $z=0$, and $\bar\rho$ is the comoving background density.  

So, one measure of the best link-length is to see which one provides 
the most universal scaling.  Figure~\ref{vfv} shows the mass functions 
in these scaled units, $\nu$, and Figure~\ref{vfvfid}, shows these curves 
divided by the same fiducial model as before.  Because we only have a 
fixed mass range in the simulations, the higher redshift outputs 
mainly probe the $\nu\gg 1$ end of the mass function.  Therefore, 
in these figures, we also show results for $z=1$.  

It is not obvious that any one link length produces more self-similar 
scalings than the others.  What is more apparent is that, whatever 
the link-length, the $z=0$ abundances appear to be offset to slightly 
larger values compared to those at higher $z$.  
This is in qualitative agreement with the spherical model, which 
predicts that halos should be increasingly dense relative to the 
background at late times, meaning that the appropriate link length 
should be smaller at late times.  By using a fixed link length, 
we will overestimate halo masses, and hence the abundance at large 
$\nu$.  

A slight variation on the appropriate self-similar scaling is to 
ignore the $z$ dependence of $\delta_{\rm sc}$.  Although this has 
no physical motivation, it is a popular choice  
\cite[e.g.][]{Jenkins:2001,Reed:2003,Warren:2006}.  We have found 
that this makes the mass function slightly less universal (the offset 
at $z=0$ is slightly more pronounced), but since we are not scaling 
the link-lengths with time in the way the spherical model suggests, 
we do not think our measurements advocate strongly for including 
the $z$-dependence of $\delta_{\rm sc}$.  

\begin{figure}
\includegraphics[width=3.25in]{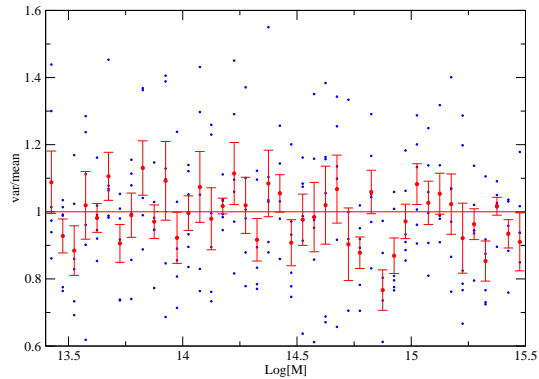}
\caption{Ratio of variance of halo counts between runs to mean halo count 
for a number of bins in mass.  
For each mass bin, error bars show the error on the mean between the six 
measurements of this ratio (the three link lengths at each of two 
redshift bins).  If the counts were Poisson, this ratio would be unity, 
with a typical spread of about $0.2$ (see text in section 3.6).  }  
\label{checkpoisson}
\end{figure}

\begin{table*}
\begin{tabular}{c c c c c c c c c c c c c c}
\multicolumn{2}{c}{Method: } & \multicolumn{2}{c}{New ML method} & \multicolumn{2}{c}{Poisson ML method}  & \multicolumn{2}{c}{$\chi^2$ method} 
& \multicolumn{2}{c}{New ML method} & \multicolumn{2}{c}{Poisson ML method}  & \multicolumn{2}{c}{$\chi^2$ method}\\
z  &    $l_{\rm link}$  &   q   &    p  &    q  &      p  &     q   &    p  &  rms(q) &  rms(p) &  rms(q) &  rms(p)  &  rms(q) &  rms(p) \\
\hline
0.0  &   0.15  &  0.82  &  0.289 &  0.805 &  0.297 &  0.803 &  0.298 &  0.008 &  0.004 &  0.007 &  0.003 &  0.006 &  0.003 \\
0.0  &   0.168 &  0.773 &  0.272 &  0.756 &  0.282 &  0.753 &  0.284 &  0.008 &  0.004 &  0.006 &  0.003 &  0.006 &  0.003 \\
0.0  &   0.2   &  0.709 &  0.248 &  0.689 &  0.26  &  0.687 &  0.261 &  0.007 &  0.004 &  0.005 &  0.003 &  0.005 &  0.003 \\
\hline
0.5  &   0.15  &  0.842 &  0.288 &  0.836 &  0.293 &  0.833 &  0.296 &  0.01  &  0.006 &  0.007 &  0.004 &  0.007 &  0.004 \\
0.5  &   0.168 &  0.792 &  0.269 &  0.784 &  0.276 &  0.785 &  0.275 &  0.009 &  0.006 &  0.006 &  0.004 &  0.006 &  0.004 \\
0.5  &   0.2   &  0.724 &  0.241 &  0.714 &  0.251 &  0.708 &  0.257 &  0.008 &  0.006 &  0.006 &  0.004 &  0.006 &  0.004 \\
\hline
\end{tabular}
\caption{Best fit parameters from three ways of fitting equation~(\ref{vfvst})
       to the halo abundances in the simulations, and the rms dispersion 
       between the 49 simulations.}
\label{massfitSTtable}
\end{table*}

\subsection{Fitting the mass function}
We fit the halo catalog to a given parametric model of the halo mass 
function in three ways, and we do this for the functional forms 
given by \cite{ST99} and \cite{Warren:2006}. In both cases 
\begin{equation}
\nu f(\nu) = \frac{m}{\rho}\,\frac{dn(m)}{d\ln m}\,\frac{d\ln m}{d\ln\nu}
\end{equation}
The first case has 
\begin{equation}
\nu f_{\rm ST}(\nu) = 
A_p\, \Bigl[1 + (q\nu)^{-p}\Bigr]\sqrt{\frac{q\nu}{2\pi}}\,\exp(-q\nu/2)
\label{vfvst}
\end{equation}
where $A_p = [1 + 2^{-p}\,\Gamma(1/2-p)/\Gamma(1/2)]^{-1}$ is chosen so 
that the integral of $f$ over all $\nu$ is unity.  This functional 
form has two free parameters, $(q,p)$.  
The second, 
\begin{equation}
\nu f_{\rm W}(\nu) = A\, \Bigl[1 + b\,(c\nu)^{-a}\Bigr]\,\exp(-c\nu/2),
\label{vfvw+}
\end{equation}
has four free parameters, because there is no requirement that the 
integral over all $\nu$ equal unity (indeed, it diverges!). 

\begin{figure}
\includegraphics[width=3.5in]{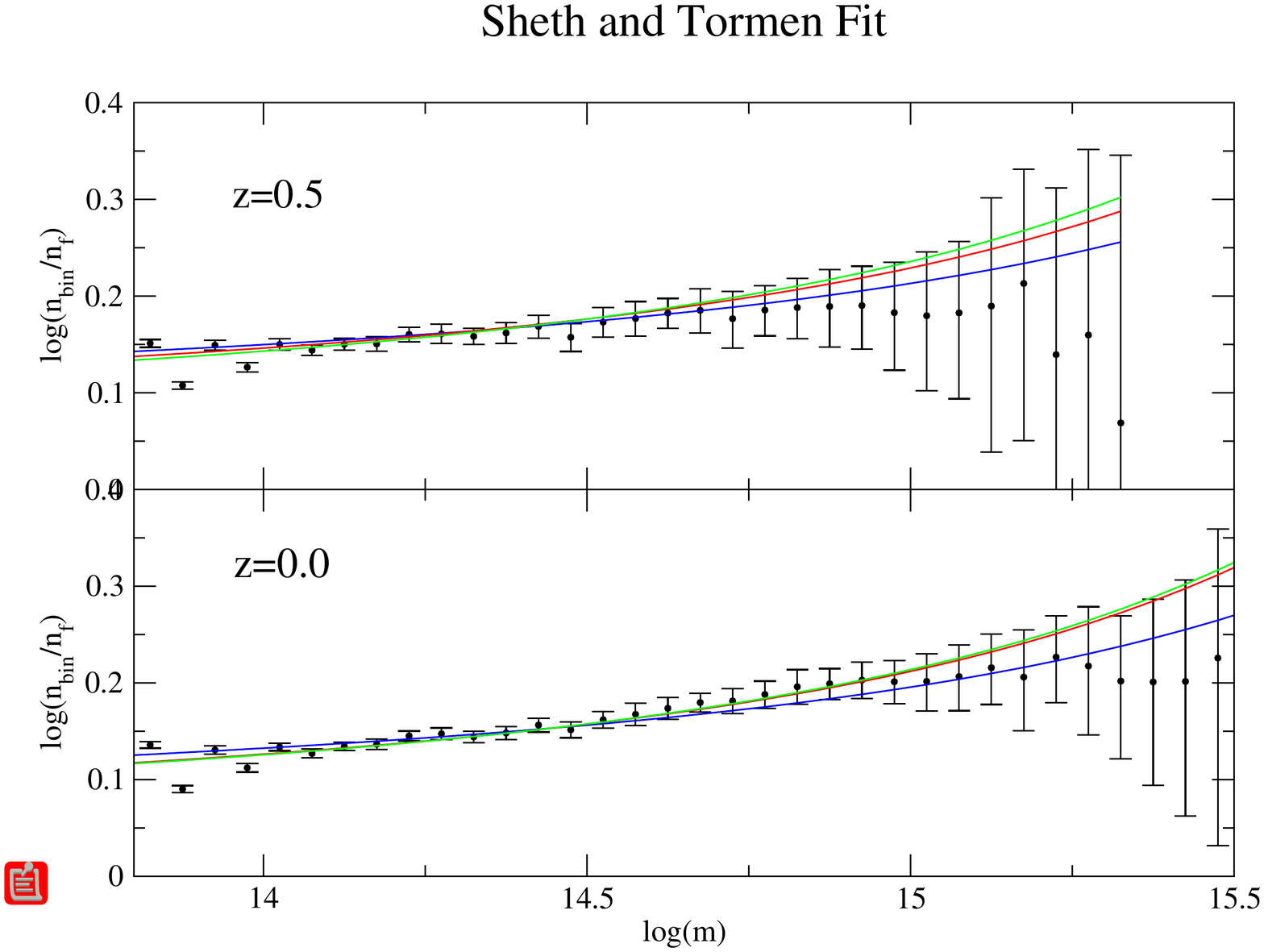}
\includegraphics[width=3.5in]{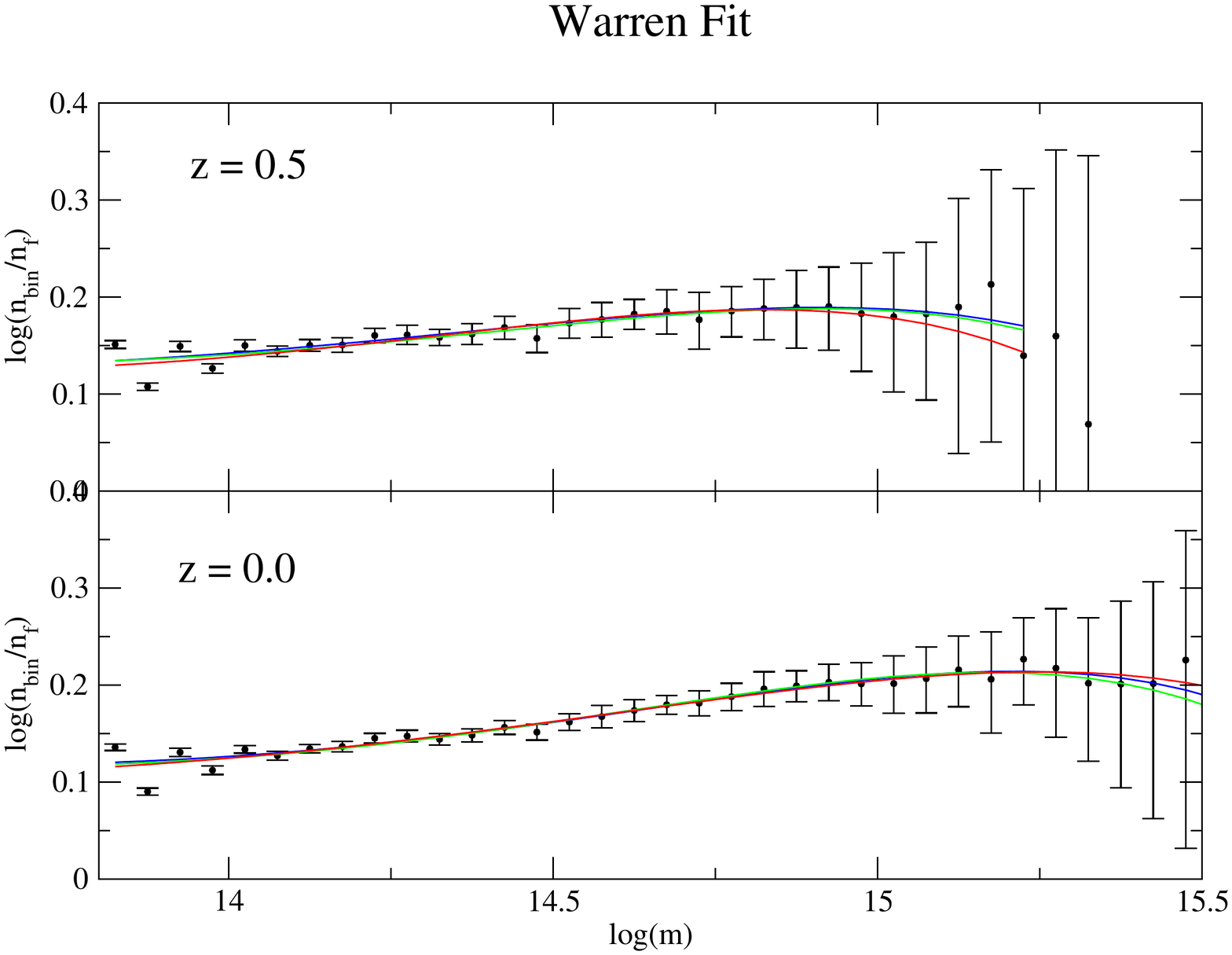}
\caption{Mass functions when the link length is 0.2, divided by a 
fiducial curve; three curves show fits to equation~\ref{vfvst} 
and~\ref{vfvw+} returned by our three algorithms:  
$\chi^2$-fit (green), Poisson ML fit (red), and new ML fit (blue). Error bars show rms between variation between simulations.}
\label{massfuncfits}
\end{figure}

Of our three fitting methods two are standard and one is new.
The two standard methods compare the theoretical model with a 
binned halo mass function, and both assume Poisson counts in a bin. 
But, whereas one approach computes a simple chi-square of the 
difference between the expected and measured counts in bins 
\citep[e.g.][]{Jenkins:2001,Reed:2007}, 
the other uses a Maximum Likelihood approach 
\citep{Warren:2006}. These methods are slightly less than ideal, 
because there is some art in choosing the size of the bin. 
In the Appendix, we describe our new method, which is a Maximum 
Likelihood estimator that does {\em not} work with binned counts.  

Since the Poisson assumption is an important ingredient in the 
first two methods (our new method makes an equivalent assumption), 
it is important to check if this assumption is accurate.  
Figure~\ref{checkpoisson} shows the ratio of the variance between runs 
to the mean count (determined by averaging over all the runs) in each bin.  
If the counts are truly Poisson, then this ratio should be unity, 
with a typical spread of about $\sqrt{2/(N-1)}$, where $N$ is the 
number of runs from which the mean and variance were estimated 
(this assumes $N\gg 1$ is large).  
The Figure shows that the Poisson assumption is good, although 
there is a hint that the variance drops below the Poisson value 
for the most massive halos.  

To minimize systematic effects due to the finite mass resolution of 
the simulation we only fit the mass function for halos with more 
than 105 particles:  i.e., $M\simeq 6.3 10^{13} h^{-1}M_\odot$. 
For the two fitting methods that require binned counts, the bin 
widths were 0.05~dex, except for the highest mass bin, which was 
enlarged to include at least 80 halos (in most cases this last bin 
contains more than 200 halos). For each bin, the rms of the 49 
simulations was used as a weight when performing the chi-square fit. 
Figure~\ref{massfuncfits} shows the results; all three estimators 
return similar fits to the measurements.  

In practice, when fitting to equation~(\ref{vfvst}), the best-fit 
$p$ and $q$ values vary little from one simulation to another, 
so if one averages $p$ and $q$ over the 49 runs, then the mass 
function associated with these averaged values is a good description 
of the average measured mass function.  
Table~\ref{massfitSTtable} shows the mean and rms dispersion of 
$p$ and $q$, derived from averaging the best fit values for each of 
the 49 simulations.  

The uncertainties in $p$ and $q$ are correlated.  
We argue in the Appendix that this may be understood, at least for 
our new estimator, in terms of the mass fraction that is predicted to 
lie above our minimum mass threshold \citep[following][]{Sheth:2003}.  
This quantity is very well measured in each simulation and, for the 
case of equation~(\ref{vfvst}), this means that the best fit $p$ 
and $q$ are expected to lie along a simple well-defined curve, and 
they do.)    

Reporting our results of fitting to equation~(\ref{vfvw+}) is 
less straightforward.  This is because this functional form has 
four free parameters, so two other measured quantities are 
required for tracking correlation between parameters.  The most 
natural candidates are the mean and mean square mass of the halos 
that are above threshold.  
These constraints give rise to a complicated set of islands in 
parameter space, thus compromising any attempt to describe the 
uncertainty range on the best fit parameters in terms of simple 
lower and upper limits.  (I.e., if one rises slightly above the 
level of the global minimum, one includes many other local minima.)  
In this case, the curves we show are for the parameters obtained by 
combining the halo catalogs from all the individual simulations, and 
then performing the fit.  Figure~\ref{fitW} illustrates.  Notice 
that the parameter $c$ is rather well constrained, whereas the other 
two are not.  This is because we are essentially only fitting the 
high mass end, where the counts are falling exponentially and 
the parameters $a$ and $b$ matter little.  Indeed, whereas the 
various best-fit parameter combinations all produce essentially the 
same counts at the lowest masses we probe, they differ (slightly) only 
at high masses.  

\begin{figure*}
\begin{tabular}{cc}
\includegraphics[width=3.5in]{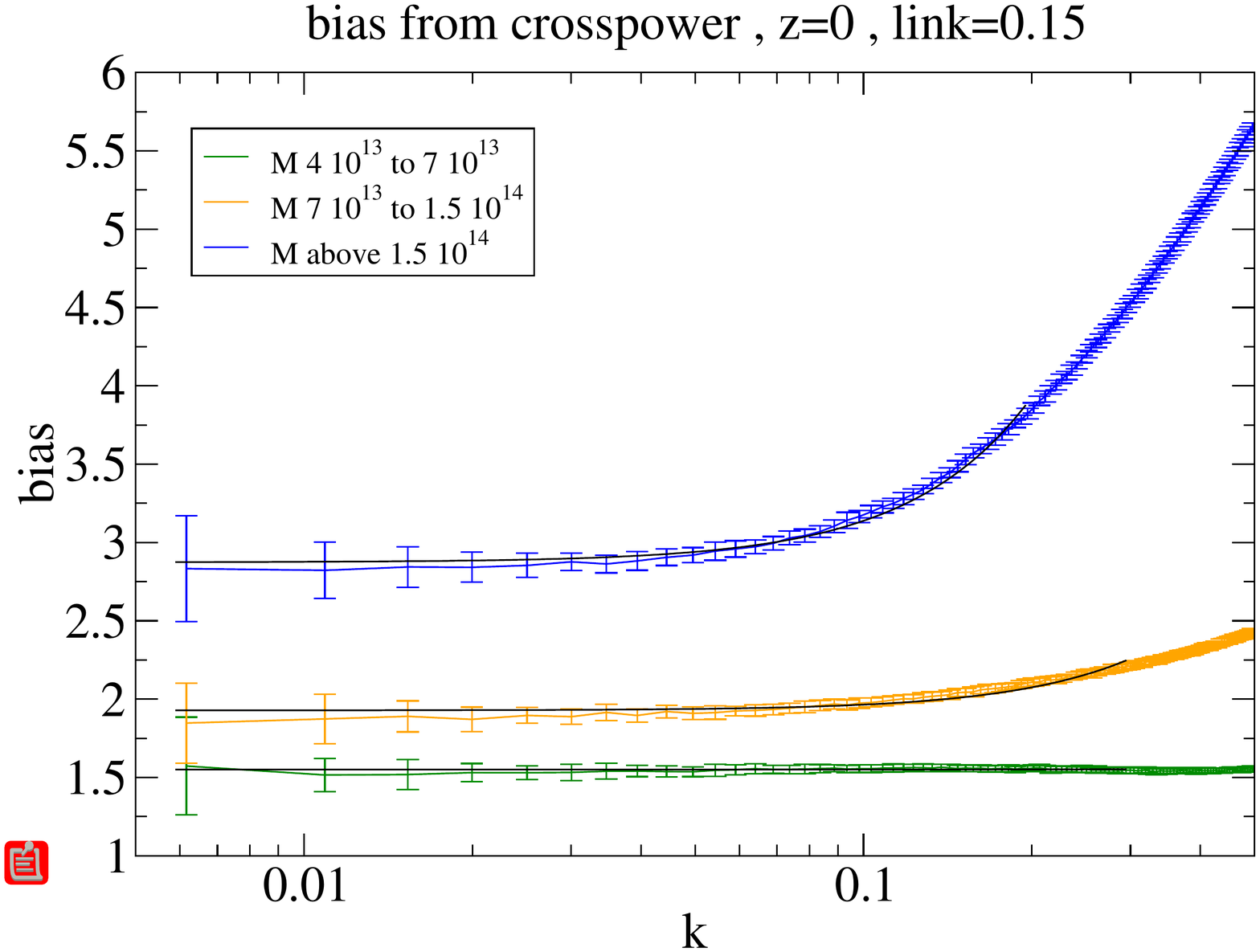} &
\includegraphics[width=3.5in]{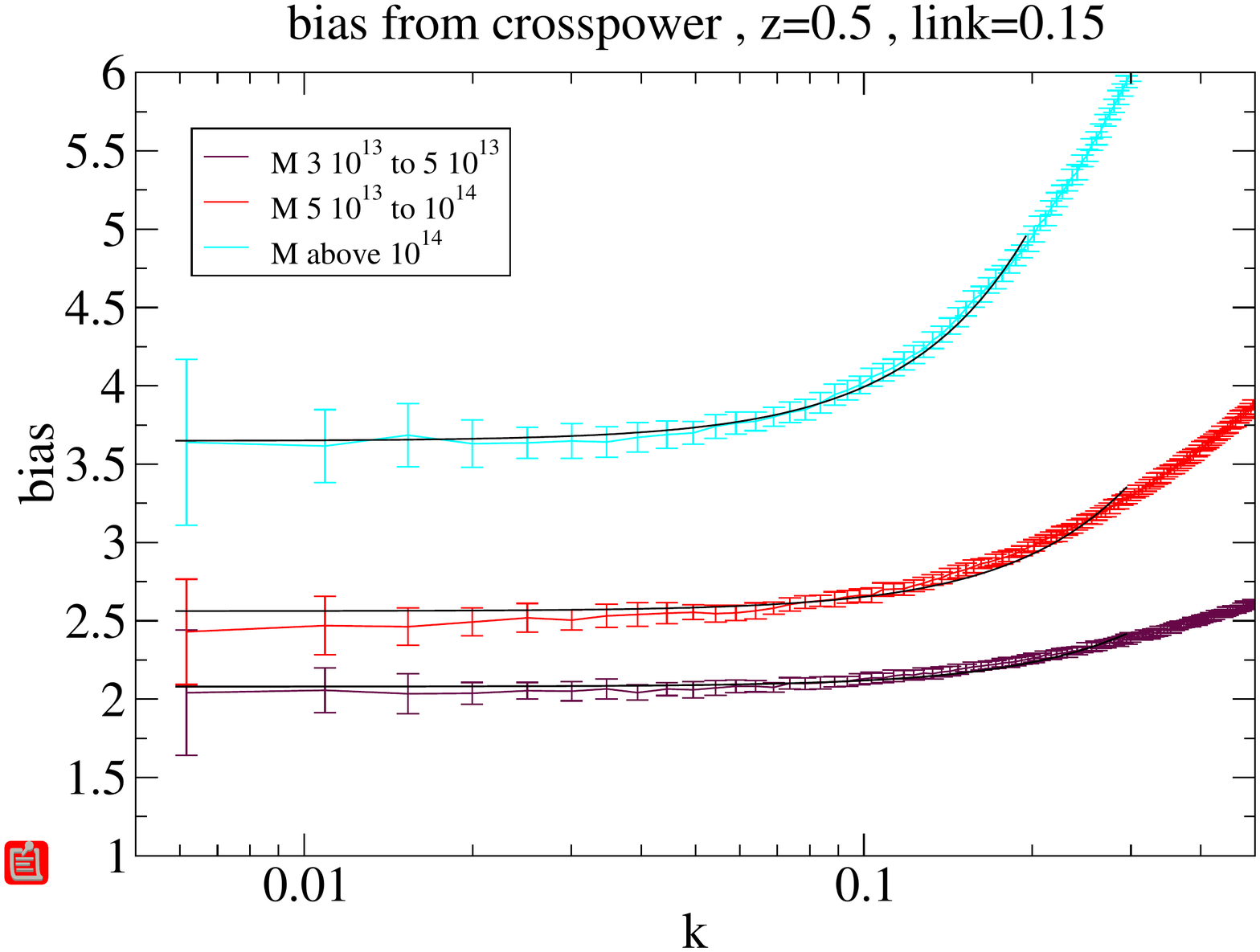} \\
\includegraphics[width=3.5in]{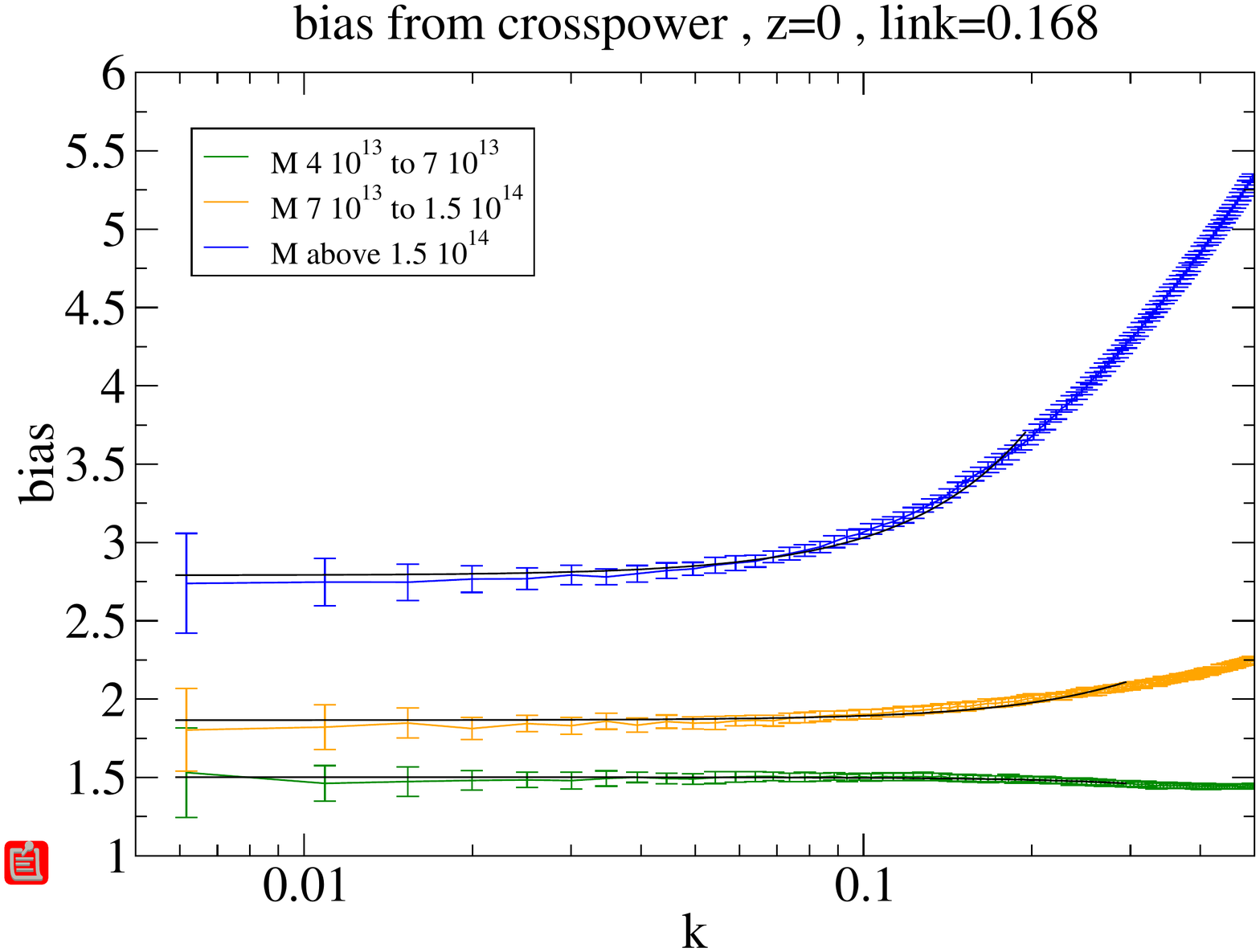} &
\includegraphics[width=3.5in]{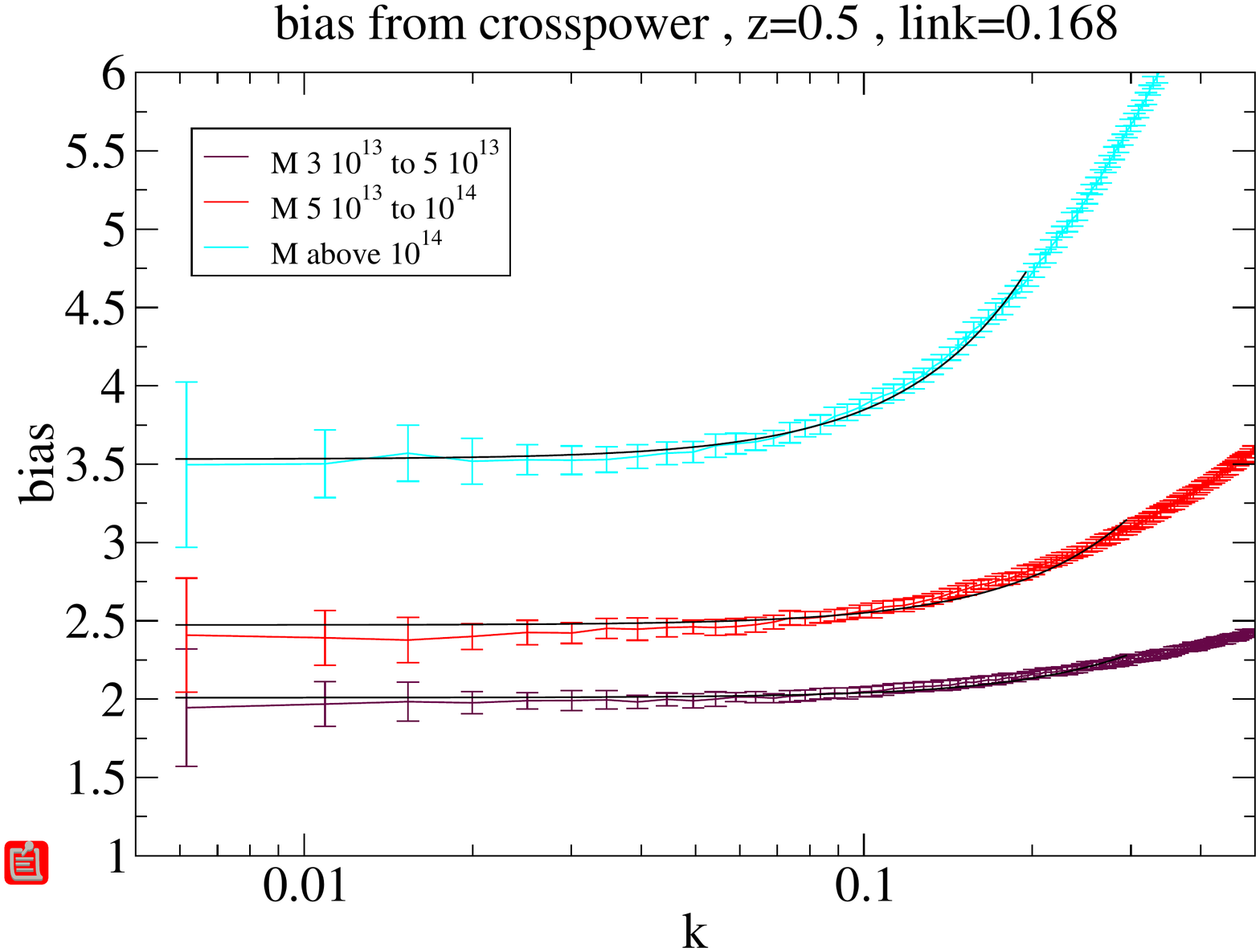} \\
\includegraphics[width=3.5in]{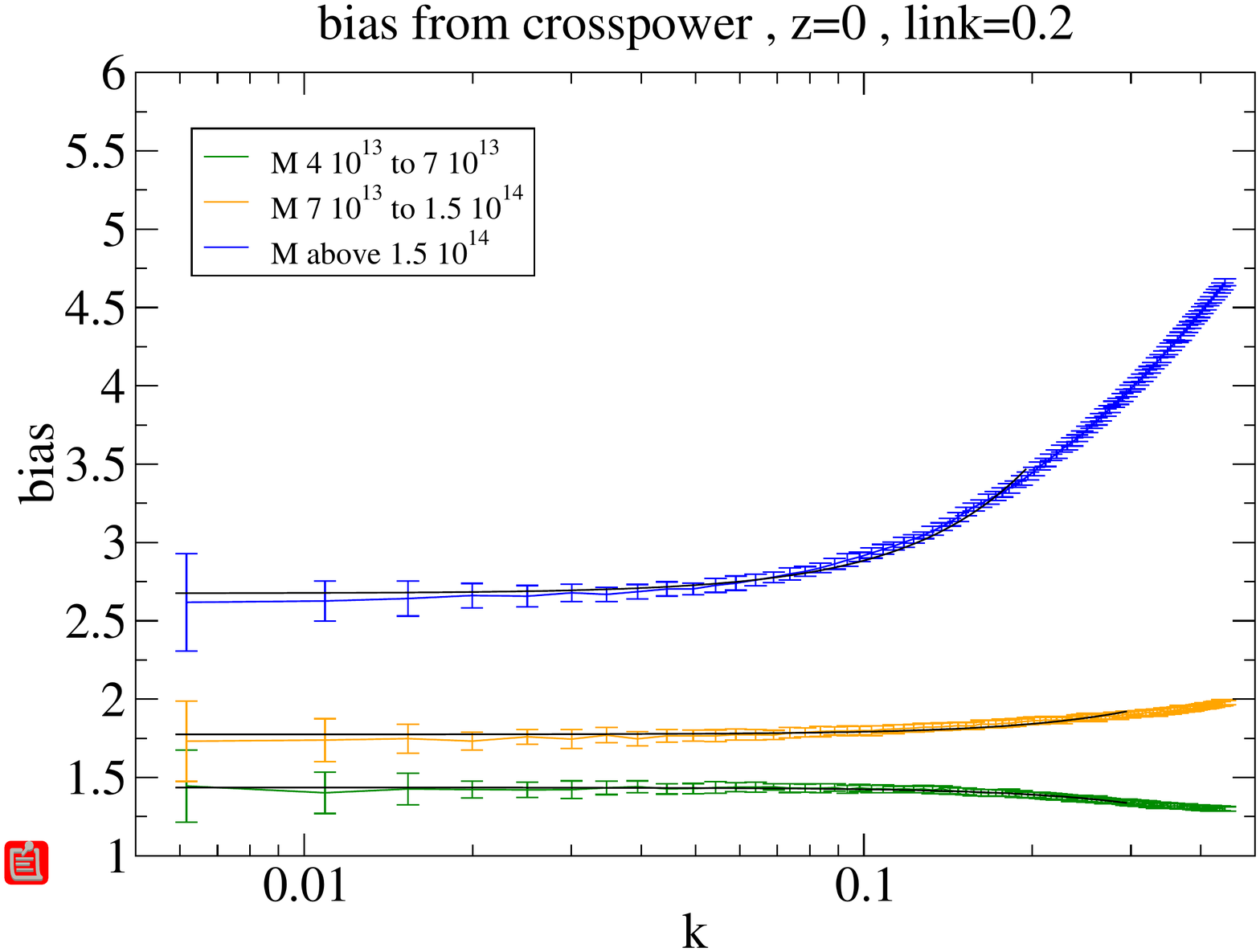} &
\includegraphics[width=3.5in]{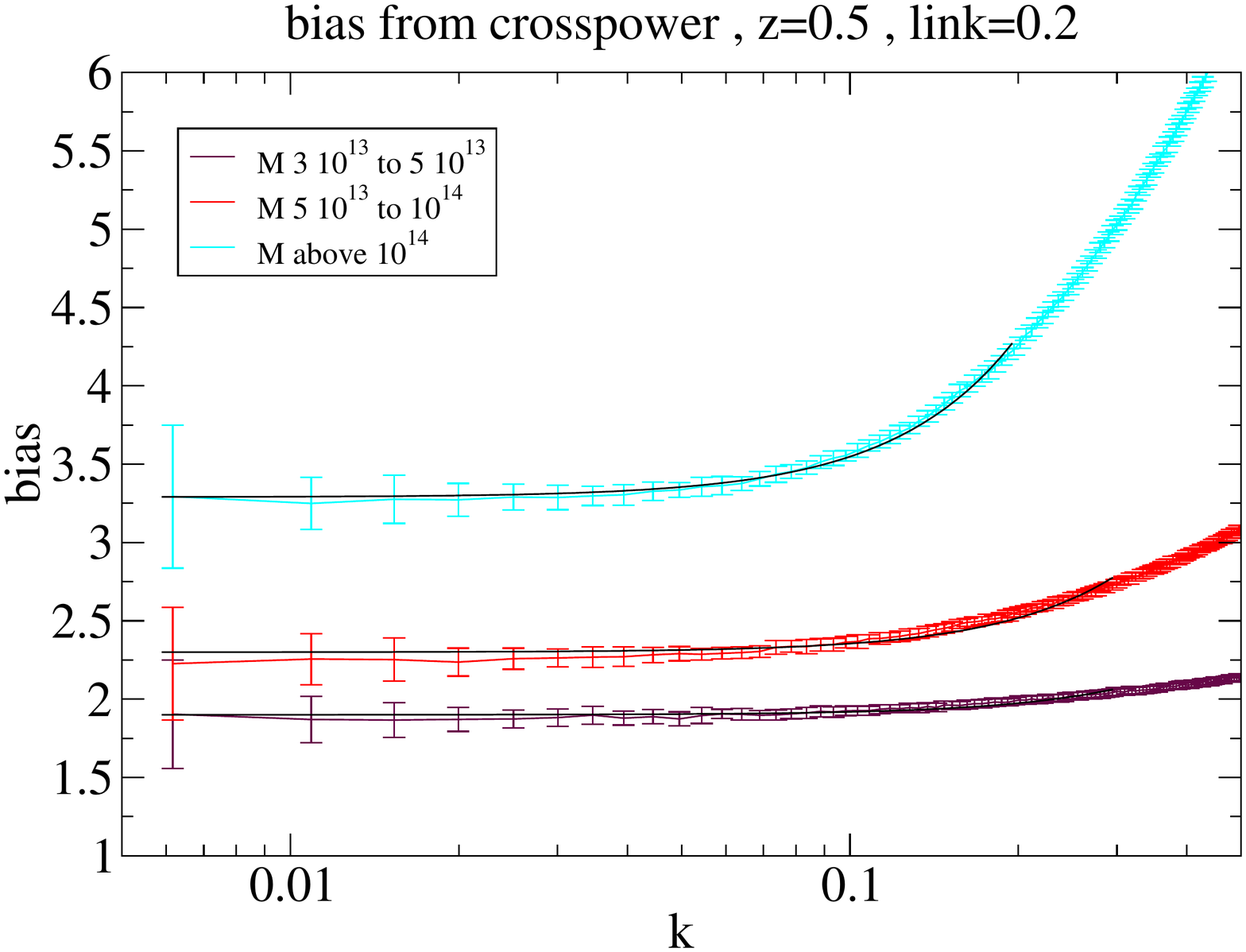} \\
\end{tabular}
\caption{Halo-mass bias from cross power spectra. Left panels show results 
       at $z=0$; right panels at $z=0.5$. 
       From top to bottom, linking lengths are 0.15, 0.168 and 0.2. Error bars show rms variation between simulations. Black solid lines are fits to the k dependence of bias between $k=[0.006,0.2]$ for the highest mass bins and $k=[0.006,0.3]$ for the other mass bins.}
\label{bias}
\end{figure*}

Before concluding this section, it is worth noting that, for a 
given link-length, the value of $p$ changes little with $z$.
In contrast, for a fixed $z$, the value of $p$ decreases systematically 
as $l_{\rm link}$ increases, suggesting that the intuitively appealing 
notion of the set of particles linked together by longer link-lengths 
at an earlier time being the same as the set linked together by a 
shorter link-length at a later time, is not correct in detail.  

\subsection{Halo-mass cross power-spectra}
For the reasons discussed earlier, we have measured the halo-mass 
cross power spectra for all our halo catalogs, and so obtained the 
large scale bias for different halo mass bins. 

\begin{table*}
\begin{tabular}{c c c c c c c c c c c c c}
\multicolumn{3}{r}{Mass range: } & & \multicolumn{3}{c}{Low} & \multicolumn{3}{c}{Medium}  & \multicolumn{3}{c}{High}\\
$z$  &    $l_{\rm link}$  &   $q$   &    $p$  &    $b_1$  &     $ b_2$  &     $b_3$   &    $b_1$  &  $b_2$ &  $b_3$ &  $b_1$ &  $b_2$  & $b_3$ \\
\hline
0.  &	0.15	& 0.82	& 0.289	& 1.6	& -0.2589  &  -1.422 &	1.914 &	0.1515	& -3.134 & 2.728 & 2.468 & -6.795 \\
0.  &	0.168	& 0.773	& 0.272	& 1.534	& -0.3326  &  -1.111 &	1.83  &	0.01092	& -2.675 & 2.616 & 2.094 & -6.378 \\
0.  &	0.2	& 0.709	& 0.248	& 1.442	& -0.4203  &  -0.7046 &	1.715 &	-0.1604	& -2.061 & 2.461 & 1.619 & -5.734 \\
\hline
0.5 &	0.15	& 0.842	& 0.288	& 2.079	& 0.4327  &  -4.113	& 2.481	& 1.385	& -6.54	 & 3.435 & 5.32 & -8.319 \\ 
0.5 &	0.168	& 0.792	& 0.269	& 1.982	& 0.24	  & -3.556	& 2.361	& 1.056	& -5.868  & 3.28  & 4.598 & -8.311 \\
0.5 &	0.2	& 0.724	& 0.241	& 1.847	& 0.003493 &  -2.801	& 2.196	& 0.6481 & -4.912 & 3.066 & 3.679 & -8.028 \\ 
\hline
\end{tabular}
\caption{Peak-background split bias factors (Appendix~\ref{pbsk} gives 
   explicit expressions) with the free parameters $p$ and $q$ obtained 
   from using our new ML method to fit the halo abundances to 
   equation~(\ref{vfvst}) (see Table~\ref{massfitSTtable}).} 
\label{biasfrommeanpq}
\end{table*}

\begin{table}
\begin{tabular}{c c r r c c c c }
$z$  & $l_{\rm link}$ &  $M_{\rm min}$  &   $M_{\rm max}$  &    bias  &  rms & $b_{\nu}$ & $b_{\zeta}$\\
\hline
0.0 & 0.15 &   4   &  7   &  1.53  & 0.05  & 1.55 & 0.02  \\
0.0 & 0.15 &   7   & 15  &  1.89  & 0.05  & 1.93 & 3.67  \\
0.0 & 0.15 &  15   & $10^{5}$  &  2.88   & 0.06  & 2.87 & 26.4  \\
0.5 & 0.15 &   3   &  5  &  2.05  & 0.06  & 2.08 & 3.88  \\
0.5 & 0.15 &   5   & 10  &  2.50  & 0.06  & 2.56 & 9.10  \\
0.5 & 0.15 &  10           & $10^{5}$  &  3.64  & 0.11  & 3.64 & 35.5  \\
\hline
0.0 & 0.168 &  4   &  7  &  1.48  & 0.05  & 1.50 & -0.45 \\
0.0 & 0.168 &  7   & 15  &  1.83  & 0.05  & 1.87 & 2.81  \\
0.0 & 0.168 &  15  & $10^{5}$  &  2.79  & 0.06  & 2.79 & 24.1  \\
0.5 & 0.168 &  3   & 5   &  1.99  & 0.06  & 2.01 & 3.11  \\
0.5 & 0.168 &  5   & 10  &  2.42  & 0.07  & 2.47 & 7.73  \\
0.5 & 0.168 &  10           & $10^{5}$ &  3.52  & 0.09  & 3.53 & 31.5  \\
\hline
0.0 & 0.2   &  4    &  7  &  1.42  & 0.06  & 1.43 & -1.13 \\
0.0 & 0.2   &  7    & 15  &  1.74  & 0.06  & 1.77 & 1.69  \\
0.0 & 0.2   &  15   & $10^{5}$ &  2.67 &  0.06 & 2.68 & 20.9  \\
0.5 & 0.2   &  3    & 5  & 1.88  & 0.06  & 1.90 & 1.86  \\
0.5 & 0.2   &  5    & 10 &  2.26 &  0.06 & 2.30 & 5.46  \\
0.5 & 0.2   &  10   & $10^{5}$ &  3.28 &  0.08 & 3.29 & 25.8  \\
\hline
\end{tabular}
\caption{{\small Large-scale bias for three bins in halo mass.  
     Halo masses are in units of $10^{13} h^{-1} M_{\sun}$. 
     The bias was measured from the halo-mass cross spectrum at 
     $k=0.03 ~h/{\rm Mpc}$, while the parameters $b_\nu$ and $b_\zeta$ 
     are a fit to the scale dependence of the bias between 
     $k=[0.006,0.2]$ for the high mass bin and $k=[0.006,0.3]$ 
     for the other two mass bins.}
        }
\label{pkbtable}
\end{table}

Figure~\ref{bias} shows the ratio of $P_{\rm hm}$ to the power spectrum 
of the mass at $z=0$ for three bins in halo mass. The three panels 
show results for the three linking lengths. In all cases, for $k$ 
below $0.05h^{-1}$Mpc, the bias is approximately independent of $k$. 
\cite[The strong $k$-dependence at larger $k$ is consistent with 
previous work, e.g.,][]{ST99}.  
This large scale bias is largest for the halo catalog from the 
shortest linking length.  This is not surprising, since the bias is 
expected to increase with halo mass, and a halo of a given mass with 
this length will only be more massive when the link length is longer. 
Thus, for example, halos at the high end of the middle mass bin may 
have been in the larger mass bin when the link length was longer. 
Their stronger clustering increases the bias for the small 
link-length catalogs.  

If we had found that the longer link-length halo catalogs from an 
earlier time were essentially the same as the shorter link-length 
catalogs at a later time, then we would be able to use the continuity 
equation to relate the bias of the high-$z$ long-$l_{\rm link}$ 
objects to the bias of the low-$z$ short-$l_{\rm link}$ objects.  
Although not exact, this should still give a good qualitative idea 
of the bias:  $(b_z-1) = (b_0-1)(D_0/D_z)$ so, for $b_1>1$, we expect 
the high-$z$ sample to have a larger bias factor.

\subsection{Relation to peaks bias}
In view of our discussion of peaks bias, we have fitted our 
measurements to functions of the form $b_\nu + b_\zeta k^2$.
These parameters are reported in Table~\ref{pkbtable} 
together with the value of the bias at $k=0.03h^{-1}$Mpc and its 
rms error.  In most cases, the quadratic form is not a good fit to 
the $k$-dependent bias at $k>0.2h$Mpc$^{-1}$ -- the $k$-dependence 
is weaker.  However, Table~\ref{pkbtable} shows that the amplitude 
of the quadratic piece increases rapidly as $m$ increases, in 
qualitative agreement with expectations.

We have found that the radii $R_{pk}$ required to match the values 
of $b_\nu$ and $b_\zeta$ in the large scale $\nu$ limit 
(equation 38 in \cite*{Desjacques2008}) are about $8-9h^{-1}$Mpc 
for the largest mass bin, and smaller for the other bins.  
These radii are comparable to the initial Lagrangian radii of the 
halos, so they are not unreasonable.  However, to see if the scaling 
with mass is quantitatively correct, we should account more carefully 
for how the range in halo masses maps to that in peak smoothing scales, 
as well as for the effects of nonlinear evolution on $b_\nu$ and 
$b_\zeta$.  This is beyond the scope of our paper.  

\subsection{Comparison with predicted large-scale bias}
We are now in a position to compare the measured large scale bias 
factor with that predicted from fitting the mass function and 
applying the peak background split to estimate $b_1$.  
The peak-background split prediction is 
\begin{equation}
\label{pbspred}
b_1 \equiv 1 - \frac{\partial\ln {\rm d}n/{\rm d}m}{\partial\delta_c},
\end{equation}
so $b_1$ associated with equations~(\ref{vfvst}) and~(\ref{vfvw+}) is 
\begin{eqnarray}
\label{b1st}
b_1^{ST} &=& 1 + \frac{q\nu -1}{\delta_c} + \frac{2 p/\delta_c}{1 + (q\nu )^p} 
 \qquad {\rm and}\\
b_1^{W} &=& 1 + \frac{c \nu -1}{\delta_c} 
       + \frac{2 a b + b + (c \nu)^a}{\delta_c\, (b + (c\nu)^a)}.
\label{b1w+}
\end{eqnarray}

\begin{figure}
\vspace{-1cm}
\includegraphics[width=3.6in]{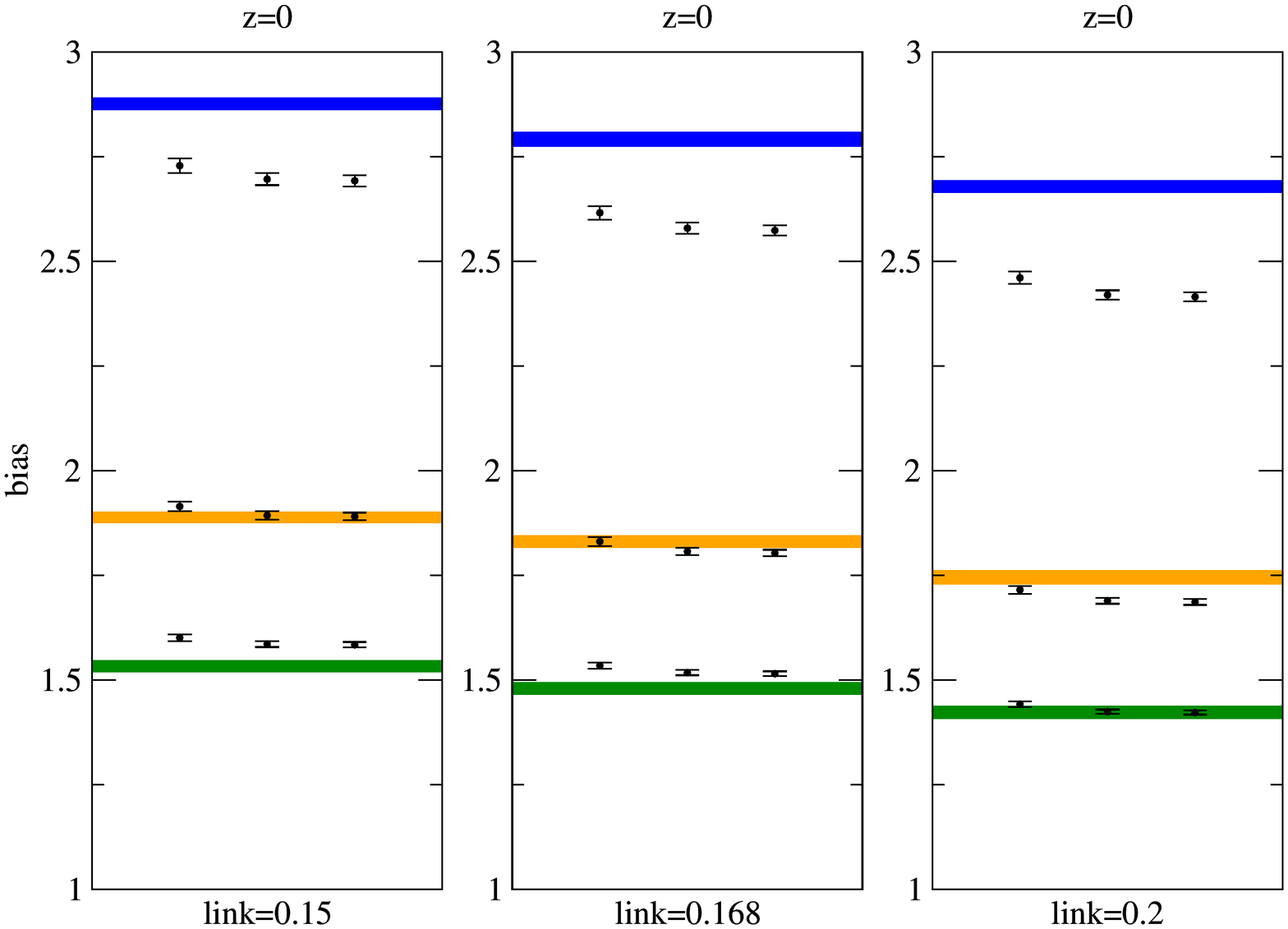}\\
\vspace{-1cm}
\includegraphics[width=3.6in]{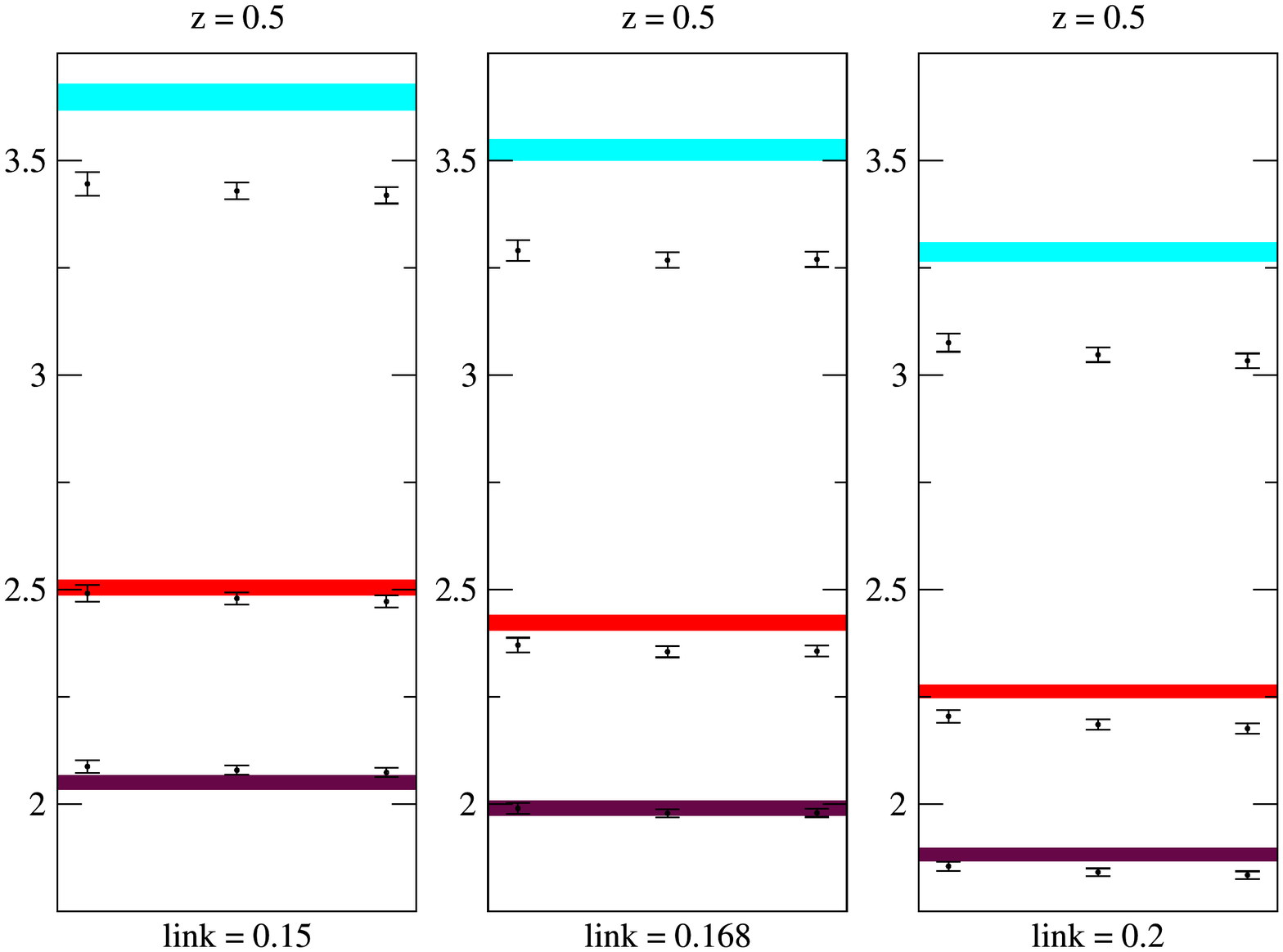}\\
\caption{Comparison of measured large scale bias factor (thick solid line)
  with the predicted $b_1$ of equation~(\ref{b1st}), for the same three 
  bins in halo mass shown in the previous figure (higher masses have 
  larger bias factors).  The parameters $p$ and $q$ of $b_1$ are 
  obtained from fitting the mass function to equation~(\ref{vfvst}).  
  For each mass bin, the three symbols with error bars show the 
  predictions associated with our three ways of fitting the mass 
  function; the error bars show the scatter in the bias between 
  the 49 simulations, divided by $\sqrt{49}$. 
  Upper panel shows results at $z=0$, lower panel at $z=0.5$.}
\label{comparebst}
\end{figure}

The thick solid lines in Figure \ref{comparebst} show the measurement, 
$P_{\rm hm}/P_{\rm mm}$ at $k=0.03~h$Mpc$^{-1}$. 
The thickness of the lines shows the two-$\sigma$ range for the 
measurement, i.e., two times the error on the mean value. 
Each triple of symbols shows the predicted bias ($b_1$ of 
equation~\ref{b1st}) associated with our three ways of fitting the 
mass function to equation~(\ref{vfvst}).  
Clearly, they give similar results. The error bars show the scatter 
in the predicted peak-background split bias between the 49 simulations 
(i.e., we use the best fit $p$ and $q$ obtained from fitting the halo 
abundances in a simultion to predict its $b_1$; the scatter in $p$ and 
$q$ between simulations translates into scatter in $b_1$).  
The upper and lower panels show results at $z=0$ and $z=0.5$ respectively.

\begin{figure}
\vspace{-1cm}
\includegraphics[width=3.6in]{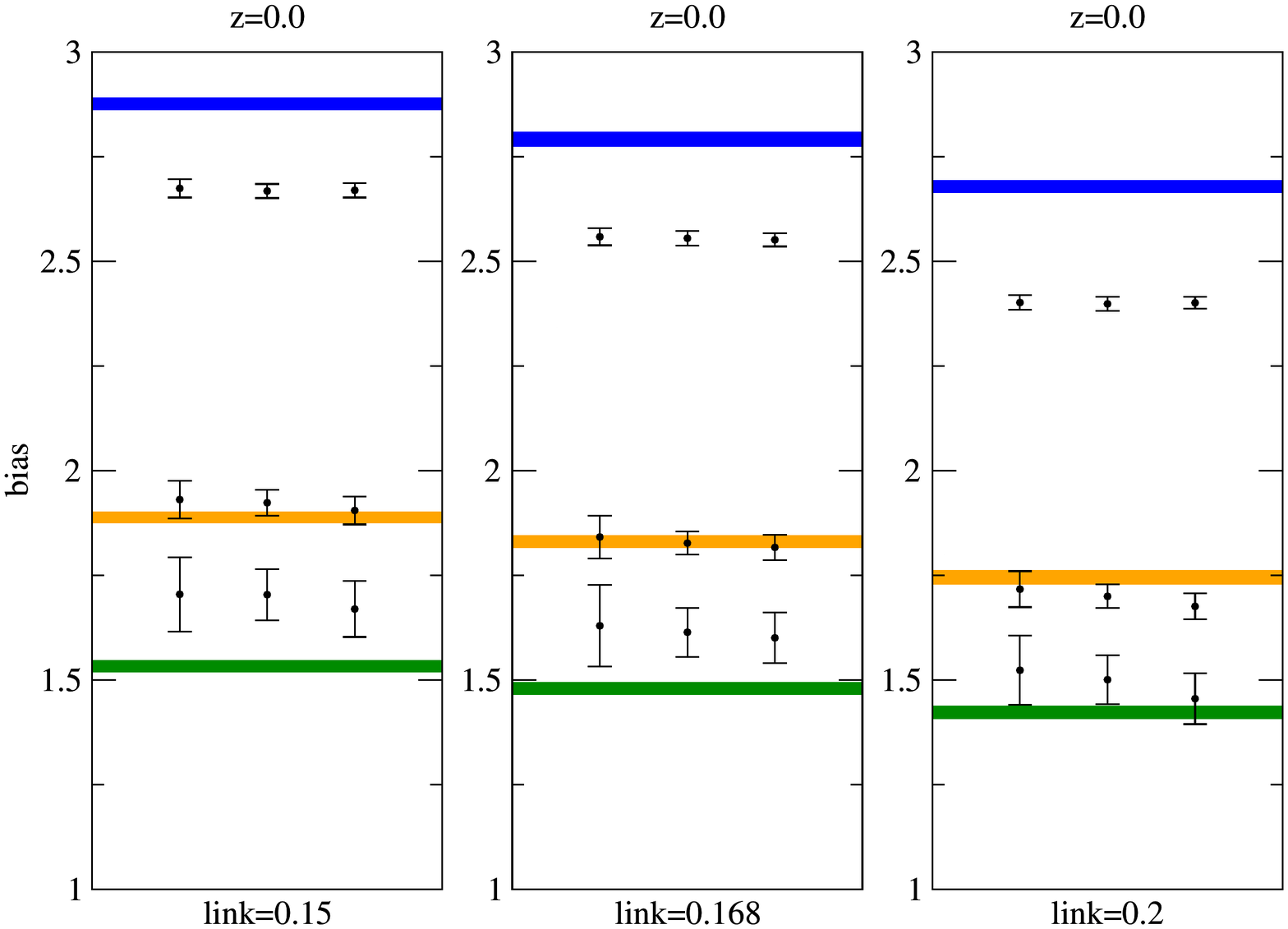}\\
\vspace{-1cm}
\includegraphics[width=3.6in]{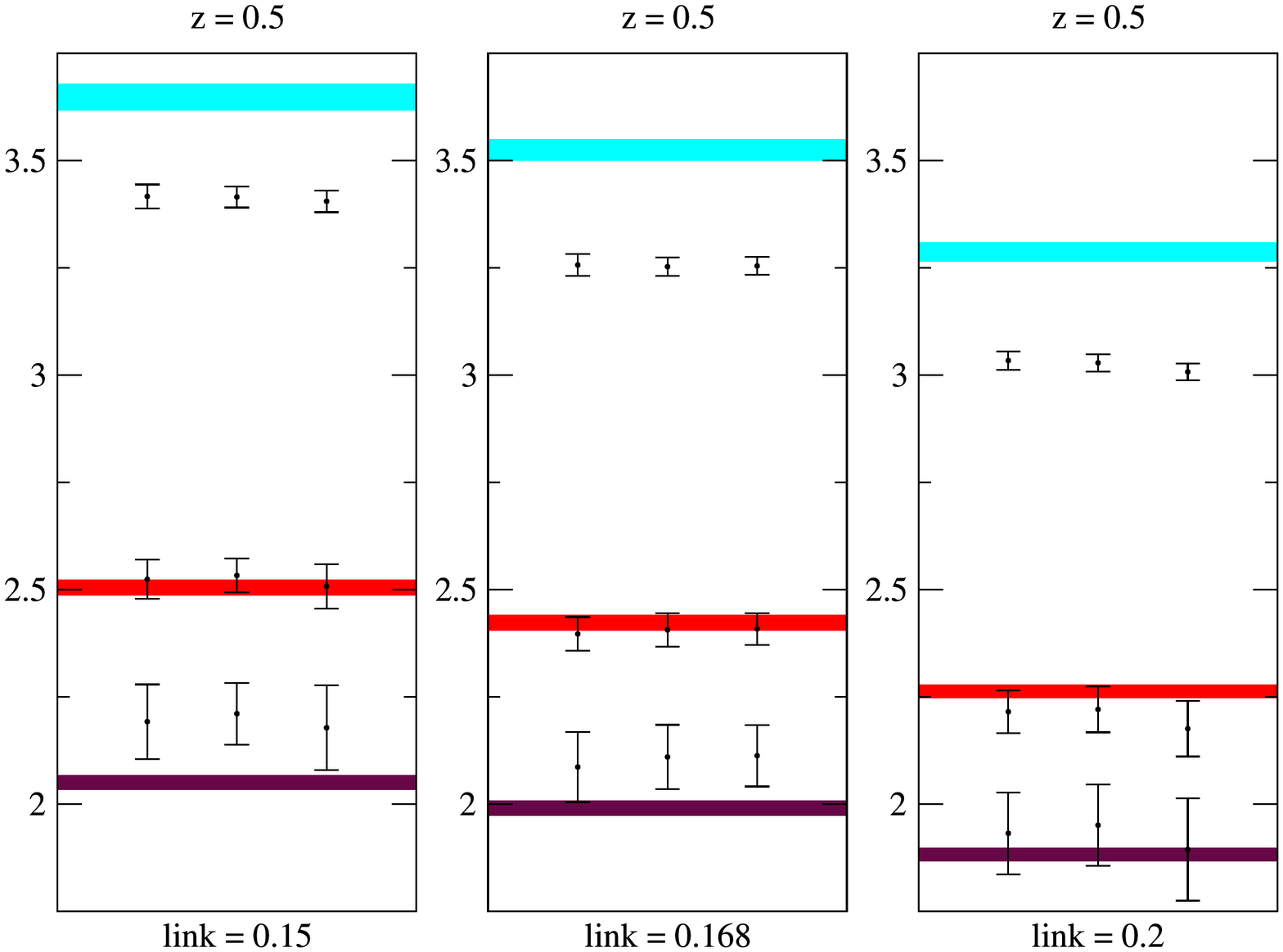}\\
\caption{Same as previous figure, but now $b_1$ is from 
        equation~(\ref{b1w+}), with parameters from fitting the mass 
        function to equation~(\ref{vfvw+}).}
\label{comparebw+}
\end{figure}

The differences between the measurements and the predicted values 
of $b_1$ are statistically significant, especially for masses which 
are large compared to $M_*$.  Figure~\ref{comparebw+} 
shows that this is {\em not} due to the parametric form assumed 
for the halo mass function: fitting to equation~(\ref{vfvw+}) and 
using the associated expression for $b_1$ (equation~\ref{b1w+}), 
yields similar results.  (There is one obvious difference:  at high 
masses, the uncertainty on the predicted $b_1$ is similar to that 
associated with equation~\ref{b1st}, but at lower masses, the 
uncertainty associated with equation~\ref{b1w+} is substantially 
larger.  This is because, at high masses, both formulae for $b_1$ 
are sensitive only to the scale of the exponential cut-off in halo 
counts, which is determined by the parameters $q$ and $c$ respectively.  
At lower masses, the other parameters also matter, of which there are 
more for equations~\ref{vfvw+} and~\ref{b1w+} than for 
equations~\ref{vfvst} and~\ref{b1st}.)  
We find qualitatively similar effects for all our choices of 
$l_{\rm link}$.  

What should we make of the discrepancy between the measured large 
scale bias and $b_1$ at high masses?  Following the discussion of 
Section~\ref{bnl}, such differences are not unexpected, because the 
peak-background split bias relation is nonlinear.  
As a result, the expected large scale bias factor $b_\times$ depends 
on the higher order bias parameters $b_2$ and $b_3$ as well as 
$b_1$ (see equation~\ref{nonlinearbias}).  Like $b_1$, these also 
depend on halo mass, and the parametrization of the halo mass function.  
Explicit formulae are provided in Appendix~\ref{pbsk}, and
Table~\ref{biasfrommeanpq} provides the numerical values associated 
with the fits to equation~(\ref{vfvst}).

\begin{figure*}
\begin{tabular}{cc}
\includegraphics[width=3.5in]{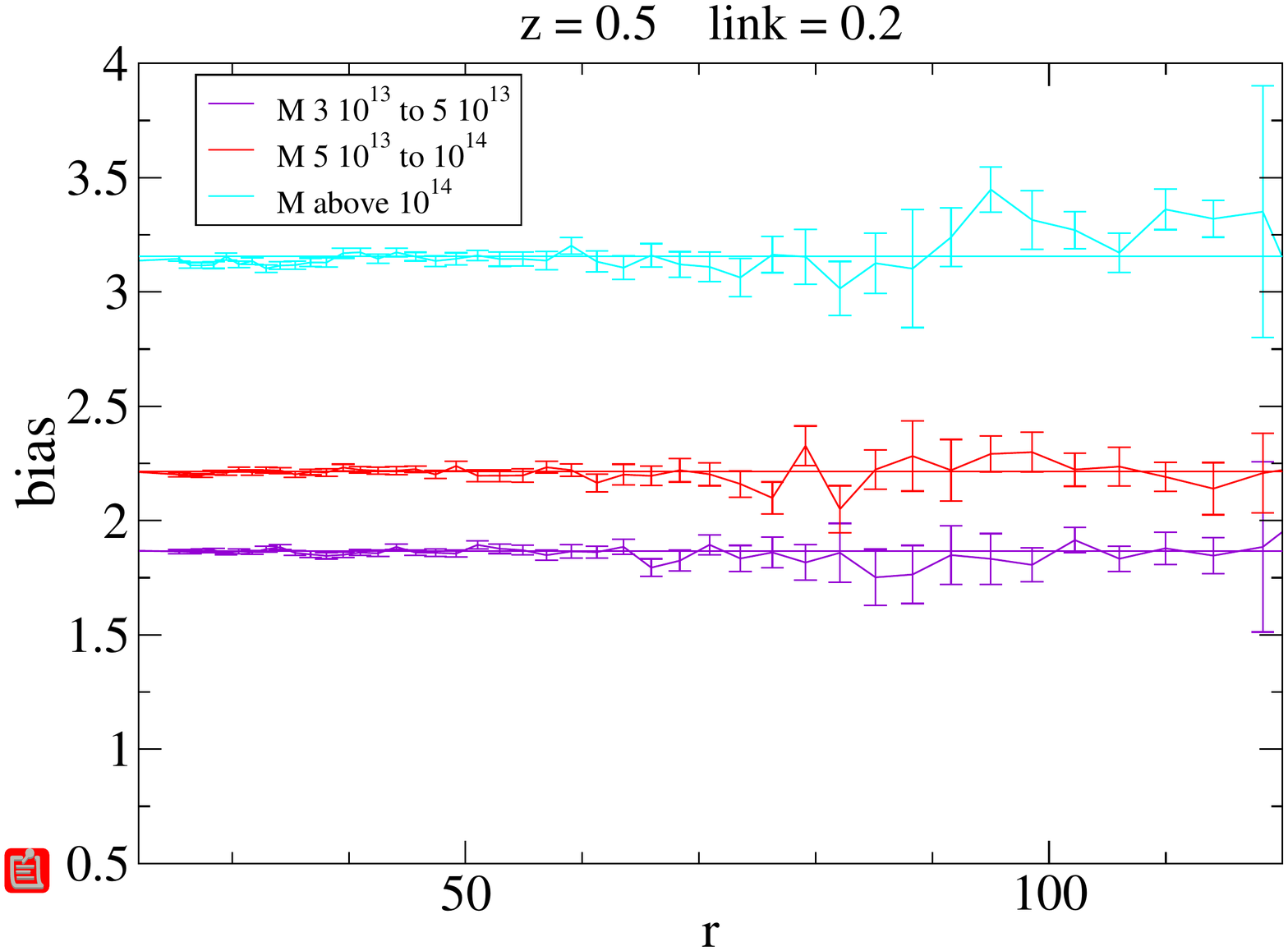} &
\includegraphics[width=3.5in]{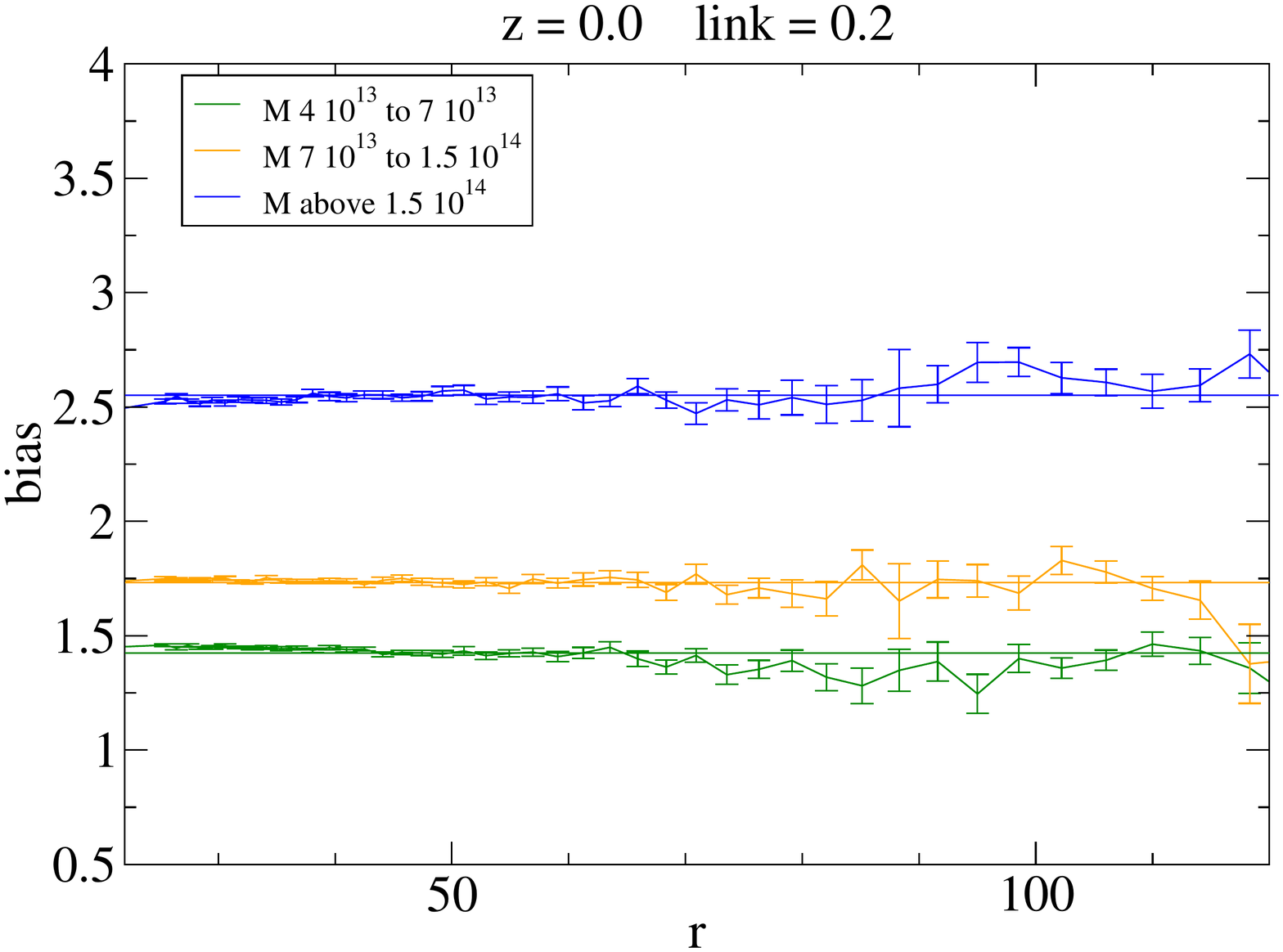} \\
\end{tabular}
\caption{Configuration space estimate of halo bias,  
       $\sqrt{\xi_{hh}/\xi_{dm}}$, for the same mass bins 
       as in previous Figures, when $l_{\rm link}=0.2$ at   
       $z=0.5$ (left) and $z=0$ (right). Error bars show the
error on the mean value betweeen simulations.}
\label{biasfromxi}
\end{figure*}

Unfortunately, the expected difference depends on a smoothing 
scale $R$ for which we have no underlying theory.  On the other 
hand, equation~(\ref{bcross}) shows that we expect 
$b_\times\approx b_1$ for our lower mass bins, but that 
$b_\times \ge b_1$ at very large masses, in qualitative agreement 
with our measurements.  (For lower masses than we are studying 
here, we expect $b_\times\le b_1$.)
Therefore, we have treated $R$ as a free parameter, to allow 
equation~(\ref{nonlinearbias}) for $b_\times$ to fit as well 
as possible.  The predicted difference between $b_\times$ and 
$b_1$ which results sometimes has the wrong sign, because $b_3$ 
can be large and negative (see Table~\ref{biasfrommeanpq}).  
The differences at large masses are qualitatively consistent with 
our measurements if we ignore higher order terms in $\sigma^4$ and 
we set $b_3=0$, although there is no theoretical justification for 
either of these steps.  And if we do this, then we are unable to 
match the measurements at lower masses.  
Thus, while equation~(\ref{nonlinearbias}) can sometimes account 
qualitatively for the differences seen in Figures~\ref{comparebst} 
and~\ref{comparebw+} ($b_2$ and $b_3$ are both negative 
in the low-mass limit), it cannot account in detail for the observed 
differences.  This suggests that the deterministic nonlinear local 
bias model does not provide a sufficiently accurate description of halo 
bias.

\subsection{Comparison with bias from configuration space}

So far we have been measuring the large scale bias from simulations 
in Fourier space using $P_{hm}$.  But one can also measure it in 
configuration space from the correlation function $\xi_{hh}/\xi_{dm}$. 
Figure~\ref{biasfromxi} shows $\sqrt{\xi_{hh}/\xi_{dm}}$ for the same 
three halo mass bins when $l_{\rm link}=0.2$.  Error bars show the 
error on the mean value between simulations.  A constant bias is a 
good description of the measurement on scales between $25-75h^{-1}$Mpc. 
The average value of this ratio, computed between 
$r = [40,60]h^{-1}$Mpc, is shown by the solid horizontal lines.  
At scales close to the acoustic peak (105$h^{-1}$Mpc for our 
cosmological model) the bias has some scale dependence, particularly 
for the highest mass halos, which we discuss shortly.  

Figure~\ref{comparebiasfromxi} compares the Fourier space measurement 
of $P_{hm}/P_{mm}$ (bars on the left of each panel), with the mean and 
dispersion of $\sqrt{\xi_{hh}/\xi_{dm}}$ (thick solid bars on right of 
each panel).  (Recall that, for each simulation, these ratios are 
averaged over the range $r=[40,60]h^{-1}$Mpc.)  The widths of the bars 
show the $2\sigma$ error on the mean measured bias (i.e, the rms 
dispersion times $2/\sqrt{49}$), indicating that these two measures of 
the bias are slightly but significantly different for the highest mass 
bin.  Each pair of error bars shows the two peak background split 
predictions for $b_1$ (equations~\ref{b1st} and~\ref{b1w+}, and recall 
that the latter has substantially larger uncertainties) for each 
of the three methods we use when fitting the mass function (from left 
to right, these are New ML, Poisson ML, $\chi^2$-method).  Notice that 
the predictions are closer to the configuration space measurement than 
the other one, but the difference is still significant.  

Unfortunately, it is not straightforward to compare $b_\xi$ of 
equation~(\ref{bauto}) with our measurements, because the theory 
calculation is for the correlation function of the smoothed halo 
field (divided by that of the similarly smoothed mass field), whereas 
our measurements of $\xi_{hh}$ and $\xi_{dm}$ are made on the 
unsmoothed point distributions.  
Nevertheless, because we measure $b_\xi < b_\times$, and this is 
qualitatively consistent with equation~(\ref{bauto}), we might ask 
what effective smoothing radius is required to explain the difference.  
For our large mass bins, this radius is of order $R\sim 40h^{-1}$Mpc.  
However, although this would make $b_\xi = b_\times$, it does not 
explain the magnitude of the difference from $b_1$.  

We noted that the halo bias has some scale dependence around the 
acoustic peak scale (105$h^{-1}$Mpc for our cosmological model).  
This scale dependent halo bias is consistent with the trends reported 
in \citep{Smith:2007, Smith:2008} that have since been confirmed by 
a number of authors \citep{Sanchez08, Sanchez09, Horizon09}.  

\subsection{Halos from spherical overdensity}\label{sectionSO}

It is well known that some objects identified by a Friends of Friends 
algorithm may have dumb-bell like shapes.  In this case, the algorithm 
labels as a single massive object what might better be classified as 
two separate objects of smaller mass.  This changes how the abundance 
{\em and} the clustering depend on mass, so one might wonder if some 
of the discrepancy with the peak-background split predictions we find 
can be attributed to our choice of group-finder.  

\begin{figure}
\includegraphics[width=3.6in]{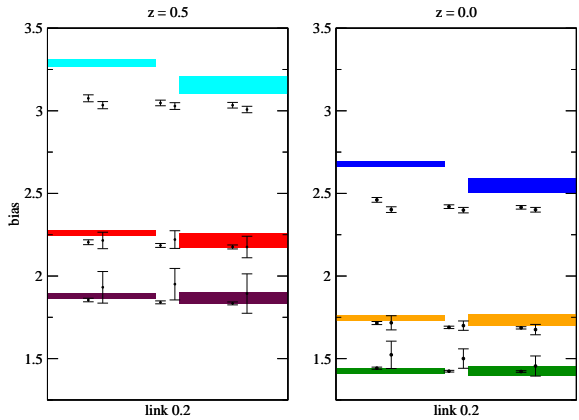}
\caption{Comparison of large scale bias estimates for the same halo mass 
  bins as in previous figures when $l_{\rm link}=0.2$.  Thick bars 
  show the measured $P_{hm}/P_{mm}$ (left) and 
  $\sqrt{\xi_{hh}/\xi_{mm}}$ (right), and symbols with error bars show 
  the linear bias parameter $b_1$ predicted from the peak background split.} 
\label{comparebiasfromxi}
\end{figure}

In this section, we perform the same analysis as before, but now 
using halos identified with a spherical overdensity (SO) requirement.  
Halos were identified as spherical regions, each 200 times denser 
than the background, in the $z=0$ outputs of our simulations by 
J. Tinker following standard methods.  
We compute the abundance, cross-power bias $b_{\times}$, and 
autocorrelation bias $b_{\xi}$ for three bins in halo mass.  
Whereas the two higher mass bins are the same as before, 
the lowest mass bin is slightly different, due to details of how 
the halo finder was run.  
Results for these measured bias factors are shown as bars in 
Figure \ref{biasplotSO}, together with the peak-background split 
prediction from their mass function (black dots with error bars).
These show that $b_{\times}$ is about 5\% larger than $b_{\xi}$, 
which is itself larger than the peak-baground split prediction.  
These are in the same sense, and have the same magnitude as our 
previous results based on FoF halos (Figure~\ref{comparebiasfromxi}).  
As an extra test we have computed $b_1$ for the higher mass bin by
fiting the mass funcion of SO halos only in the mass bin range
instead of the wider range available. In this case the difference
between $b_1$ and $b_\xi$ got reduced to half, but it remains still
significant. We conclude that our finding that $b_1\ne b_\times\ne b_\xi$ does 
not depend on how halos were identified.  


\section{Discussion and conclusions}\label{conclusions}
The peak-background split argument is commonly used to relate the 
abundances of dark matter halos to their spatial clustering. We have 
found that this estimate of the bias between halos and the dark matter 
is not accurate to better than $\sim 10$ percent when compared with 
different measures of large scale bias, particularly for the most 
massive halos.  We did not test the intermediate or low mass regime.  

\begin{figure}
\centering
\includegraphics[width=3.6in]{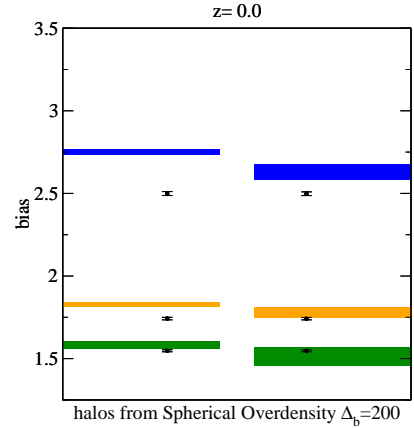}
\caption{Same as previous figure, but now for halos identified using 
  an SO algorithm.  Results are shown for the same mass bins as before, 
  except that the lowest mass bin is from $5.93\cdot 10^{13} M_{\odot}$ to $7.0\cdot10^{13} M_{\odot}$.  
  Thick bars show the measured $P_{hm}/P_{mm}$ (left) 
  and $\sqrt{\xi_{hh}/\xi_{mm}}$ (right); the thickness of the bars 
  indicates the two-$\sigma$ range. 
  Symbols with error bars show the linear bias parameter $b_1$ 
  predicted from fitting equation~(\ref{vfvst}) to the halo 
  abundances using the Poisson and $\chi^2$ methods.  
  Error bars show the rms scatter between realizations. }
\label{biasplotSO}
\end{figure}

Our results are insensitive to
a) how exactly we define halos,
b) the exact functional form of the mass function and
c) how the mass function was fitted.
We have checked this by exploring three friends-of-friends linking 
lengths for defining the halo catalogs, 0.15, 0.168 and 0.2 
(see Figures \ref{massfuncfig}--\ref{vfvfid}), as well as 
using a spherical overdensity criterion (Section~\ref{sectionSO});
two functional forms for the mass function (equations~\ref{vfvst} 
and~\ref{vfvw+}, for which the associated linear bias factors $b_1$ 
are given by equations~\ref{b1st} and~\ref{b1w+}); and three methods 
for fitting halo counts to these functional forms, one of which is new.  
The latter is a likelihood estimator that maximizes the probability 
that a randomly chosen particle belongs to a halo of specified 
mass; it does not require binned halo counts, thus removing the 
arbitrariness of the choice of bin size which is intrinsic to more 
standard methods. 

We have also studied the self-similarity of the mass function at 
different linking lengths for $z=0,0.5,1$ and find that it is 
qualitatively but not exactly self-similar (see Figure~\ref{vfvfid}). 
We have argued that this difference may be reduced by scaling the 
linking-length as a function of redshift as suggested by the 
spherical collapse model. 

Results for the different estimates of large-scale halo bias are 
shown for two different redshifts in Figures~\ref{comparebst} 
and~\ref{comparebw+}.  Although halo bias appears to be close to 
linear on large scales (Figures~\ref{bias} and~\ref{biasfromxi}), 
the bias factor $b_\xi\equiv \sqrt{\xi_{hh}/\xi_{dm}}$ one measures 
at large $r$ is different from $b_\times\equiv P_{hm}/P_{mm}$ measured 
at small $k$, and both are different from the peak-background split 
estimate of the linear bias factor $b_1$, at large masses where 
$b_1\ge 2$ (Figures~\ref{comparebiasfromxi} and~\ref{biasplotSO}).  
On the other hand, at lower masses where $b_1\approx 2$, 
$b_1\approx b_\times \approx b_\xi$ to within a few percent.  

We discussed possible explanations for the differences at large masses.  
For example, the contribution of nonlinear bias terms, $b_2$, $b_3$, 
etc., which are generic to the peak-background split argument (we 
provide explicit expressions in Appendix~\ref{pbsk}), make 
$b_1\ne b_\times\ne b_\xi$ (see equations~\ref{bcross} and~\ref{bauto}).  
However, the amplitude of these corrections depends on a parameter, 
$\sigma^2$, for which there is no underlying theory, other than 
the expectation that it is smaller than unity, but greater than zero.  
While nonlinear terms could explain the difference between
$b_\xi$ and $b_\times$, the differences between these bias factors 
and $b_1$ are consistent with our measurements only if we ignore 
terms of order $\sigma_R^4$ and higher, and we set $b_3=0$, although 
there is no theoretical justification for either of these steps.  
But then, to be self-consistent, we should use the same algorithm for 
the lower mass bins, and there, what (barely) worked for the high 
masses no longer works (because $b_2$ and $b_3$ are negative).  

\begin{figure}
\vspace{-1cm}
\includegraphics[width=0.95\hsize]{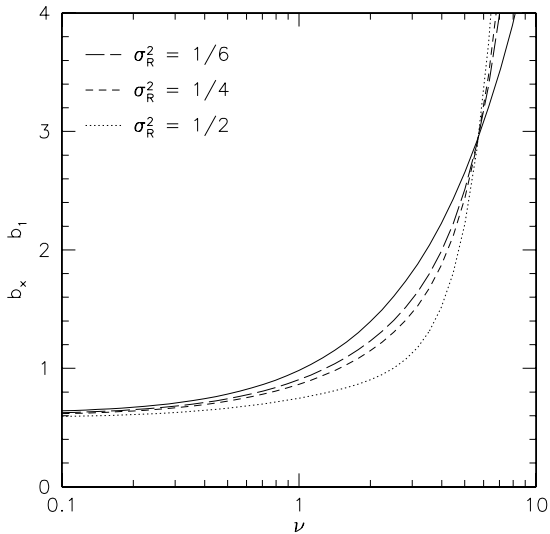}
\vspace{-2cm}
\caption{Dependence of $b_\times$ (equation~\ref{bcross}, with 
terms of order $\sigma_R^4$ and higher set to zero) on the smoothing 
parameter $\sigma_R^2$, when the bias factors $b_1, b_2$ and $b_3$ 
are given by equation~(\ref{biasSTnu}) with $(p,q)=(0.25,0.7)$.  
Solid curve shows the linear bias parameter $b_1$, which corresponds 
to the $\sigma^2\to 0$ limit of $b_\times$.  }
\label{fig:bcross}
\end{figure}

Although our analysis was restricted to massive halos, it is likely 
that our conclusions about the (in)accuracy of the peak background 
split extend to lower masses.  To illustrate, Figure~\ref{fig:bcross} 
shows how the predicted $b_\times$ differs from the linear bias factor 
$b_1$, for a number of choices of the unknown parameter $\sigma_R$.  
(To make the plot, we have ignored terms of order $\sigma_R^4$ and 
higher in equation~\ref{bcross}.)  
Note that the difference between $b_\times$ and $b_1$ is not simple:  
at high masses where $b_1\ge 2$, $b_\times>b_1$, whereas the opposite 
is true at intermediate masses, and $b_\times\approx b_1$ at very 
low masses.  In recent simulations which resolve smaller halos 
\citep[e.g.,][]{MillenniumII}, the measured large scale bias is 
indeed smaller than $b_1$, in qualitative agreement with 
Figure~\ref{fig:bcross}.  However, comparison with Fig.~10 of 
\citet{MillenniumII} shows that, at the 10\% level, the quantitative 
agreement is not good.  

We conclude that more work is needed to understand the nature of 
halo bias at the few percent level.  Our results suggest that we 
are beginning to see the limitations of the local deterministic 
bias model -- while the inclusion of higher order bias terms can 
sometimes explain the qualitative difference between $b_1$, 
$b_\times$ and $b_\xi$, it does not work quantitatively for all 
masses. 
As one alternative, we considered a peaks-bias model which is linear 
but nonlocal and scale dependent in $k$-space.  More work is needed 
before a fair quantitative comparison of this model with the measurements 
can be made, but our measurements suggest qualitative agreement.  
Another, which we are pursuing, is to study models in which the 
evolution between initial and evolved fields 
(e.g., equation~\ref{localb}) is no longer a deterministic function 
of the overdensity.

Finally, we note that our expression for the bias factor implicitly assumes that the mass function has a universal form. The fact that it is not quite universal will modify the bias factor predicted by the peak-background split \citep{ST99}, although work in progress suggests this is not enough to explain the discrepancies we have found.

\section*{Acknowledgments}
This work was partially supported by NSF AST-0607747, NASA NNG06GH21G 
and NSF AST-0908241.
We thank J.L. Tinker for identifying the SO halos in our simulations.
RKS thanks J. Bagla at HRI Allahabad, S. Mei and J. Bartlett at 
Paris 7 (Diderot), and R. Skibba and A. Pasquali at the 
Max-Planck Institut for Astronomie (Heidelberg) for their hospitality 
during the course of this work.  

\bibliographystyle{mn2e}
\bibliography{myselectionrevised}

\appendix
\section{Fitting the halo mass function}\label{newML}
This Appendix defines a Maximum likelihood estimator of the halo mass 
function that does not require binned halo counts.  The key is to 
think about the mass function in {\em exactly} the same way that 
theorists do when modeling it.  Namely, the question is not: 
How many halos are there in a certain mass bin in the simulation box? 
but,
What is the probability that a randomly chosen particle in the 
simulation box was in a halo of mass $m$?  

Let ${\rm d}n(m)\,{\rm d}m$ denote the number density of haloes of 
mass $m$.  Then the fraction of particles in such haloes is 
\begin{equation}
f(m)\,{\rm d}m = \frac{m}{\bar\rho} \,\frac{{\rm d}n(m)}{{\rm d}m}.
\end{equation}
Let $f(m|{\bm\theta})\,dm$ denote a theoretical model of this 
quantity, where ${\bm\theta}$ denotes the vector of parameters which 
specifies the model.  Then the likelihood to be maximized is 
\begin{equation}
{\cal L}({\bm\theta}) = \prod_{i=1}^{N_p} f(m_i|{\bm\theta}),
\end{equation}
where the product is over all $N_p$ particles in the simulation box.  
In practice, one only measures halos down to some minimum mass.  
This modifies the estimator above to 
\begin{equation}
{\cal L}({\bm\theta}) = 
  F(m\le M_{\rm min}|{\bm\theta})^{N_p - N_{m\ge M_{\rm min}}}\ 
  \prod_{i=1}^{N_p} f(m_i|{\bm\theta}),
\end{equation}
where 
\begin{equation}
F(m\le M_{\rm min}|{\bm\theta}) \equiv 
1 - \int_{M_{\rm min}}^\infty dm\, f(m|{\bm\theta}),
\end{equation}
and $N_{m\ge M_{\rm min}}$ is the total number of particles in 
halos above the minimum mass.  We have explicitly written this as 
unity minus the integral over massive halos to allow for the possibility 
that bound halos below some mass scale may not exist (and because 
some authors choose functional forms which lead to divergences 
when integrated over all $m$).  This way of writing the probability 
shows that it is trivial to account for this possibility.  

Now, because one has found the halos, one need not draw from the 
particle list when computing the likelihood, one can use the 
(considerably smaller!) halo catalog instead.  I.e., 
\begin{equation}
{\cal L}({\bm\theta}) = 
  F(m\le M_{\rm min}|{\bm\theta})^{N_p - N_{m\ge M_{\rm min}}}\ 
  \prod_{i=1}^{N_h} f(m_i|{\bm\theta})^{N_i},
\end{equation}
where the product is now over the $N_h$ halos in the box, 
$N_i$ is the number of particles in halo $i$, and 
\begin{equation}
N_{m\ge M_{\rm min}} = \sum_{i=1}^{N_h} N_i.
\end{equation}
The derivatives of $\ln\,{\cal L}({\bm\theta})$ with respect to the 
parameters $\theta_i$ can be done analytically, so this method is 
fast.  The second derivatives provide analytic estimates of shape 
of the likelihood surface near the minimum, and hence of the 
uncertainties on the best-fit parameters.  

In practice, the mass functions of current interest are written in 
terms of the scaled variable $\nu$.   Therefore, we scale all 
masses $m$ to $\nu$ using equation~(\ref{definenu}), and then write 
the likelihood in these scaled variables before maximizing:
\begin{equation}
{\cal L}({\bm\theta}) = 
  F(\nu\le \nu_{\rm min}|{\bm\theta})^{N_p - N_{m\ge M_{\rm min}}}\ 
  \prod_{i=1}^{N_h} f(\nu_i|{\bm\theta})^{N_i}.
\end{equation}
It is straightforward but tedious to compute the first and second 
derivatives with respect to the parameters ${\bm\theta}$.  Doing so 
gives an idea of the expected accuracy of and covariances between the 
best-fitting parameters.  However, a more intuitive demonstration 
of the covariances can be got by noting that, for large $M_{\rm min}$, 
the vast majority of particles in the simulation are not assigned to 
halos, and so the line of degeneracy is driven by requiring that the 
model always produce the observed mass fraction in halos. For example,
when fitting to equation~(\ref{vfvst}), the parameters 
$p$ and $q$ must change so as to keep 
$A_p\,[\Gamma(1/2,q\nu_{\rm min}/2,\infty)/\Gamma(1/2) + 
2^{-p}\Gamma(1/2 - p,q\nu_{\rm min}/2,\infty)/\Gamma(1/2)]$ fixed.
The solid line in Figure A1 shows this curve for halos of mass
$M > 6.31\times 10^{13} h^{-1} M_\odot$ identified with 
$l_{\rm link}=0.2$ at $z=0$, at which time the mass fraction in 
halos is 0.13 (this is the mean over all 49 simulations; the actual 
fraction varies slightly from one realization to another).  
Symbols show the best fit parameters for each of the 49 simulations.

\begin{figure}
\includegraphics[width=\hsize]{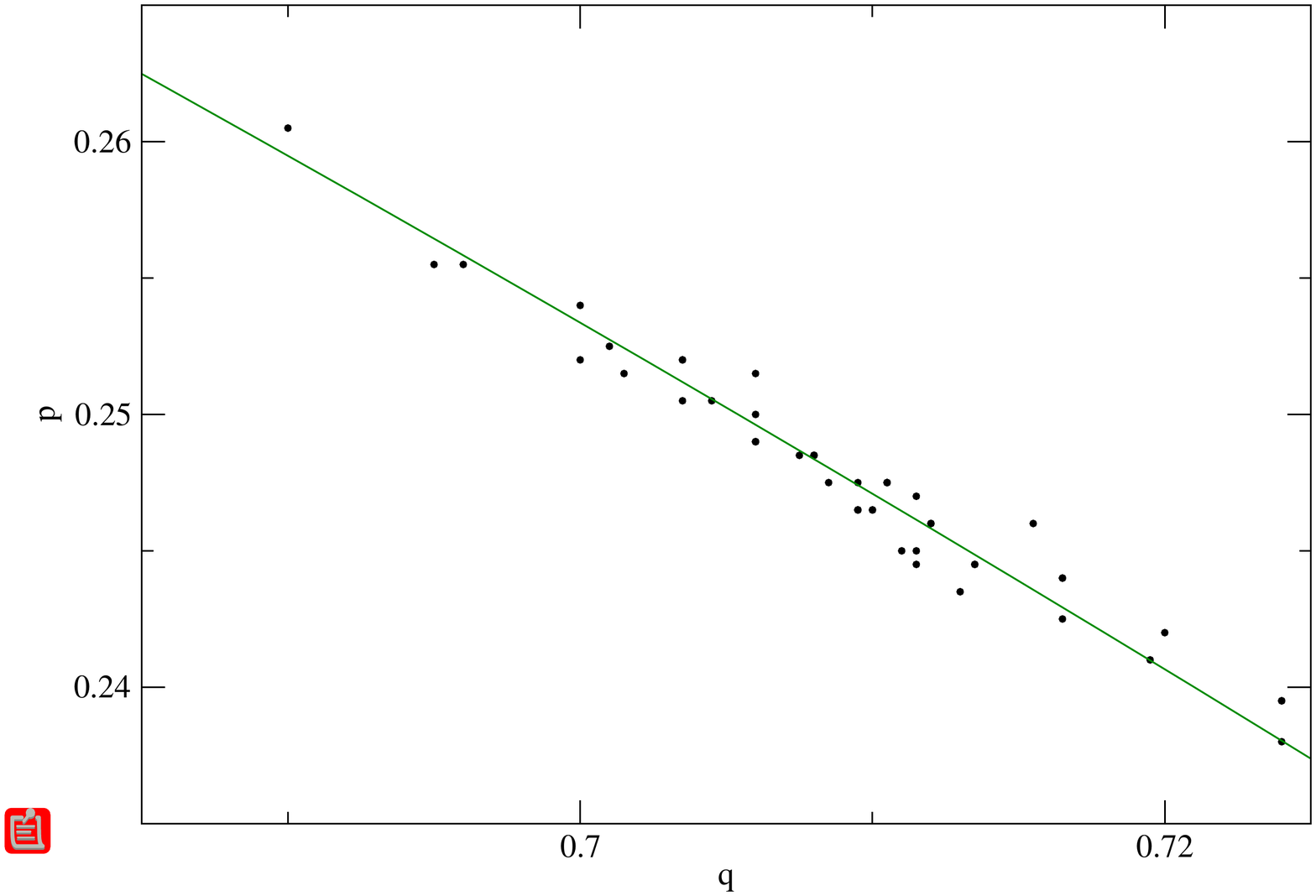}
\caption{ Best fit $p$ and $q$ parameters of equation~(\ref{vfvst}) 
 for each of the 49 simulations with linking-length 0.2 and redshift 
 $z=0$. The solid line shows a constant fraction of mass in halos 
 equal to the mean of all simulations.}
\label{plotcovar}
\end{figure}

\begin{figure}
\includegraphics[width=\hsize]{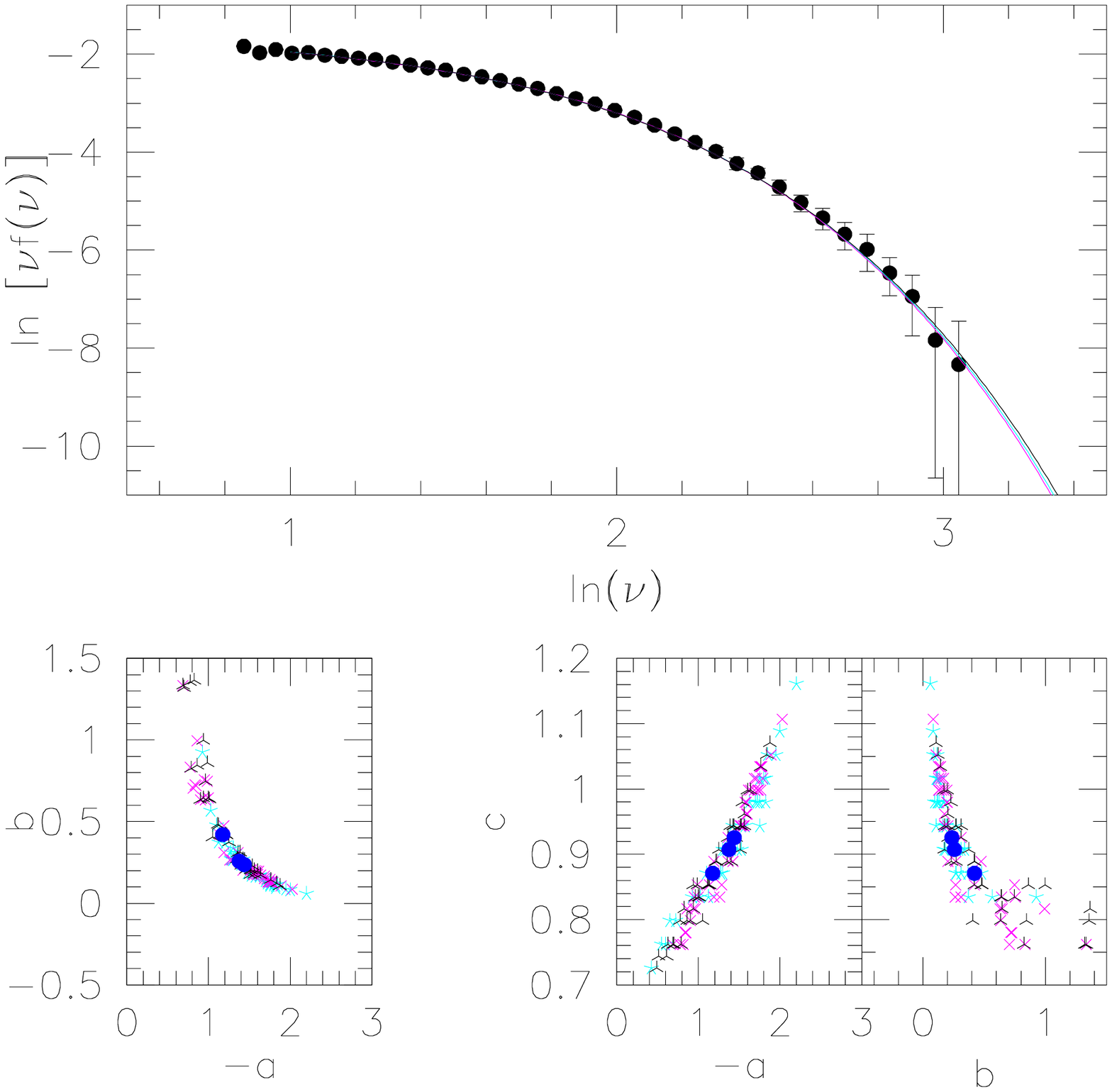}
\caption{Top:  Measured $z=0$ halo abundances (link length 0.2) 
 when the 49 simulations have been combined. Error bars show the
 rms variation between simulations. Curves show the result 
 of fitting equation~(\ref{vfvw+}) to the counts using the three methods 
 described in the main text.  All methods return essentially the 
 same counts at the lowest $\nu$ we probe; they differ slightly at 
 higher $\nu$.
 Bottom:  Covariance between best-fit parameters for each of the 
 49 simulations with linking-length 0.2 and redshift $z=0$.  
 The fractional error on $c$ is much smaller than on the other 
 parameters.  Stars, crosses and tripods show results for the 
 ML, Poisson and $\chi^2$ methods:  there is no systematic trend with 
 fitting method.  Filled solid circles show the parameters associated 
 with fitting to the combined counts.}
\label{fitW}
\end{figure}

Figure~\ref{fitW} shows a similar comparison of the measured 
covariances between best fit parameters of equation~(\ref{vfvw+}).  
We have not shown the expected correlations for this case.

\section{Bias factors}\label{pbsk}
In the peak background split ansatz, one writes the halo fluctuation 
$\delta_h$ as a power series of the mass fluctuation:
\begin{equation}
\delta_h = \sum_i \frac{b_i}{i!}\, \delta^i,
\end{equation}
and one obtains the coefficients $b_i$ by taking appropriate 
derivatives of the halo mass function, and accounting for the fact 
that halo abundances are estimated in the initial field $\delta_0$ 
rather than the evolved field $\delta$ \citep{MW:1996,MJW97,ST99}.   
Namely, one assumes there is a deterministic mapping between 
$\delta_0$ and $\delta$:  
\begin{equation}
\delta_0 = \sum_{i>0} a_i\delta^i,
\label{d0dnl}
\end{equation}
and that this mapping is given by the spherical evolution model 
\begin{equation}
a_1 = 1,\ 
a_2 = -\frac{17}{21}, \ 
a_3 = \frac{341}{567},\ {\rm and}\ 
a_4 = -\frac{55805}{130977}.
\label{sccoeffs}
\end{equation}
Then, 
\begin{eqnarray}
\label{biasSTnu}
b_1(\nu) & =& 1+\epsilon_1+E_1  \nonumber \\
b_2(\nu) & =& 2(1+a_2)(\epsilon_1+E_1)+\epsilon_2+E_2 \nonumber  \\
b_3(\nu) & =&6(a_2+a_3)(\epsilon_1+E_1)+3(1+2a_2)(\epsilon_2+E_2)\nonumber\\
           &  &  + \epsilon_3+E_3  \\
b_4(\nu) & =& 24(a_3+a_4)(\epsilon_1+E_1)+\nonumber \\
        & & + 12(a_2^2 + 2(a_2+a_3))(\epsilon_2+E_2)+\nonumber \\
        & & + 4(1+3 a_2)(\epsilon_3+E_3)+\epsilon_4+E_4 \nonumber
\end{eqnarray}
where 
\begin{eqnarray}
\epsilon_1 & = & \frac{q \nu -1}{\delta _c}, \qquad
\epsilon_2  =  \frac{q \nu  (q \nu -3)}{\delta _c^2}, \nonumber \\
\epsilon_3 & =&  \frac{q \nu  \left(q^2 \nu ^2-6 q \nu +3\right)}{\delta _c^3},
               \nonumber \\
\epsilon_4 & = & \frac{q^2 \nu ^2 \left(q^2 \nu ^2-10 q \nu +15\right)}
                    {\delta _c^4}, \nonumber\\
E_1 & = & \frac{2 p}{\delta _c (q \nu )^p+\delta _c}, \qquad
\frac{E_2}{E_1}  =  \frac{2 p+2 q \nu -1}{\delta _c}, \\
\frac{E_3}{E_1} & = &  \frac{4 p^2+6 q \nu  p+3 q^2 \nu ^2-6 q \nu -1}
                          {\delta _c^2}, \nonumber \\
\frac{E_4}{E_1} & = & \frac{2 \left(4 p^3+(8 q \nu +4) p^2 
                             +\left(6 q^2 \nu ^2-6 q \nu -1\right) p\right)}
                          {\delta_c^3}, \nonumber \\
& & \ + \frac{2( 2 q^3 \nu ^3-9 q^2 \nu ^2+q \nu -1)}{\delta_c^3} \nonumber 
\end{eqnarray}
for the mass function of equation~(\ref{vfvst}) 
\citep{Scoccimarro:2001howmany}.  

For the functional form of equation~(\ref{vfvw+}), 
\begin{eqnarray}
\epsilon_1 & = & \frac{c \nu }{\delta _c}, \qquad 
\epsilon_2  =  \frac{c \nu  (c \nu -1)}{\delta _c^2}, \qquad
\epsilon_3  =  \frac{c^2 \nu ^2 (c \nu -3)}{\delta _c^3}, \nonumber \\
\epsilon_4 & = & \frac{c^2 \nu ^2 \left(c^2 \nu ^2-6 c \nu +3\right)}
                    {\delta _c^4}, \nonumber\\
E_1 & = & \frac{2 a b}{\delta _c (c \nu )^a+b \delta _c}, \qquad
\frac{E_2}{E_1}  =  \frac{2 a+2 c \nu +1}{\delta _c},  \\
\frac{E_3}{E_1} & = &\frac{4 a^2+6 c \nu  a+6 a+3 c^2 \nu ^2+2}{\delta _c^2},
                   \nonumber \\
\frac{E_4}{E_1} & = &\frac{2 \left(4 a^3+4 (2 c \nu +3) a^2 + 
                     \left(6 c^2 \nu ^2+6 c \nu +11\right) a \right)}
                    {\delta_c^3}, \nonumber \\
& & + \frac{2(2 c^3 \nu ^3-3 c^2 \nu ^2+c \nu +3)}{\delta _c^3}. \nonumber 
\end{eqnarray}
We note that the assumption of equation~(\ref{d0dnl}) is strong, 
and only an approximation in triaxial collapse models 
\citep{Ohta04, LS08}.  Accounting for this is the subject of ongoing 
work.

\label{lastpage}

\end{document}